\documentclass[aps,pra,10pt,tightenlines,longbibliography,eqsecnum,notitlepage,amsmath,twocolumn,floatfix,superscriptaddress]{revtex4-1}
\usepackage[dvipsnames]{xcolor}
\usepackage[colorlinks=true,bookmarks=false,linkcolor=NavyBlue,urlcolor=NavyBlue,citecolor=NavyBlue,breaklinks]{hyperref}
\usepackage{amsmath}
\usepackage{amsfonts}
\usepackage{tabularx}
\usepackage{graphicx}
\usepackage{times}
\usepackage{mathtools}
\usepackage{braket}

\newcommand{\real}{\operatorname{Re}}

\newcommand{\parti}[2]{\frac{\partial #1}{\partial #2}}

\newcommand{\partit}[2]{\frac{\partial^2 #1}{\partial #2^2}}

\newcommand{\intall}{\int_{-\infty}^{\infty}}
\newcommand{\avg}[1]{\langle#1\rangle}
\newcommand{\Avg}[1]{\left\langle#1\right\rangle}

\newcommand{\abs}[1]{\left|#1\right|}
\newcommand{\bk}[1]{\left(#1\right)}
\newcommand{\Bk}[1]{\left[#1\right]}
\newcommand{\BK}[1]{\left\{#1\right\}}
\newcommand{\trace}{\operatorname{tr}}

\newcommand{\expect}{\mathbb E}

\newcommand{\norm}[1]{\lVert#1\rVert}

\newcommand{\rehacek}{\v{R}eh\'a\v{c}ek}
\newcommand{\abbrev}[1]{\textrm{#1}}

\begin{document}
\title{Resolving starlight: a quantum perspective}

\author{Mankei Tsang}
\email{mankei@nus.edu.sg}
\homepage{https://blog.nus.edu.sg/mankei/}
\affiliation{Department of Electrical and Computer Engineering,
  National University of Singapore, 4 Engineering Drive 3, Singapore
  117583}

\affiliation{Department of Physics, National University of Singapore,
  2 Science Drive 3, Singapore 117551}

\date{\today}


\begin{abstract}
  The wave-particle duality of light introduces two fundamental
  problems to imaging, namely, the diffraction limit and the photon
  shot noise. Quantum information theory can tackle them both in one
  holistic formalism: model the light as a quantum object, consider
  any quantum measurement, and pick the one that gives the best
  statistics.  While Helstrom pioneered the theory half a century ago
  and first applied it to incoherent imaging, it was not until
  recently that the approach offered a genuine surprise on the age-old
  topic by predicting a new class of superior imaging methods.  For
  the resolution of two sub-Rayleigh sources, the new methods have
  been shown theoretically and experimentally to outperform direct
  imaging and approach the true quantum limits. Recent efforts to
  generalize the theory for an arbitrary number of sources suggest
  that, despite the existence of harsh quantum limits, the
  quantum-inspired methods can still offer significant improvements
  over direct imaging for subdiffraction objects, potentially
  benefiting many applications in astronomy as well as fluorescence
  microscopy.
  
\end{abstract}

\maketitle


\section{Ingredients of the resolution problem: diffraction, photon
  shot noise, statistics}
In 1879 Lord Rayleigh proposed a criterion of resolution for
incoherent imaging in terms of two point sources \cite{rayleigh}: the
sources are said to be unresolvable if they are so close that their
images, blurred by diffraction, overlap significantly.  To quote
Feynman \cite{feynman_rayleigh}, however, ``Rayleigh's criterion is a
rough idea in the first place,'' and a better resolution can be
achieved ``if sufficiently careful measurements of the exact intensity
distribution over the diffracted image spot can be made.''  Thus
another limiting factor is the noise in the intensity measurement,
with the photon shot noise being the most fundamental source. Because
of the particle nature of light, each camera pixel can record its
energy in discrete quanta only, and ordinary light sources, including
starlight and fluorescence, introduce further randomness to the
quantum measurements \cite{mandel}.

To incorporate noise in the definition of resolution, the theory of
statistical inference offers a rigorous framework
\cite{dekker97,villiers}.  For example, a measure of resolution can be
defined in terms of parameter estimation: given a blurry and noisy
image of two point sources, how well can one estimate their separation
\cite{falconi67,tsai79,bettens,vanaert,ram}?  Or it can be framed in
terms of hypothesis testing: how well can one decide from the image
whether there is one or two sources
\cite{harris64,acuna,shahram04,shahram06}? Such statistical treatments
of resolution have garnered prominence in optical astronomy
\cite{farrell66,falconi67,lucy92,lucy92a,acuna,zmuidzinas03,feigelson}
and fluorescence microscopy \cite{ram,deschout,chao16,diezmann17,zhou19b},
where the number of photons is limited and shot noise is part of life.

\section{\label{sec_helstrom}Quantum detection 
and estimation theory}
Imaging has grown into a multidisciplinary problem that straddles
optics, quantum mechanics, statistics, and signal processing. In a
Herculean effort that began in the 1960s, Helstrom merged the subjects
into a theory of quantum detection and estimation \cite{helstrom},
which marked the beginning of quantum information theory. His aim was
to determine the best measurement, out of the infinite possibilities
offered by quantum mechanics, that optimizes the performance of an
inference task. For a given light source, the optimal performance then
represents the most fundamental limit on the resolution, valid for any
optics design that is allowed by quantum mechanics, as well as any
computational technique in data postprocessing.  In setting
fundamental limits, Helstrom's theory plays a role for sensing and
imaging not unlike the second law of thermodynamics for engines,
ruling out unphysical superresolution methods in the same manner the
second law rules out perpetual-motion machines.

\begin{figure}[htbp!]
\centerline{\includegraphics[width=0.45\textwidth]{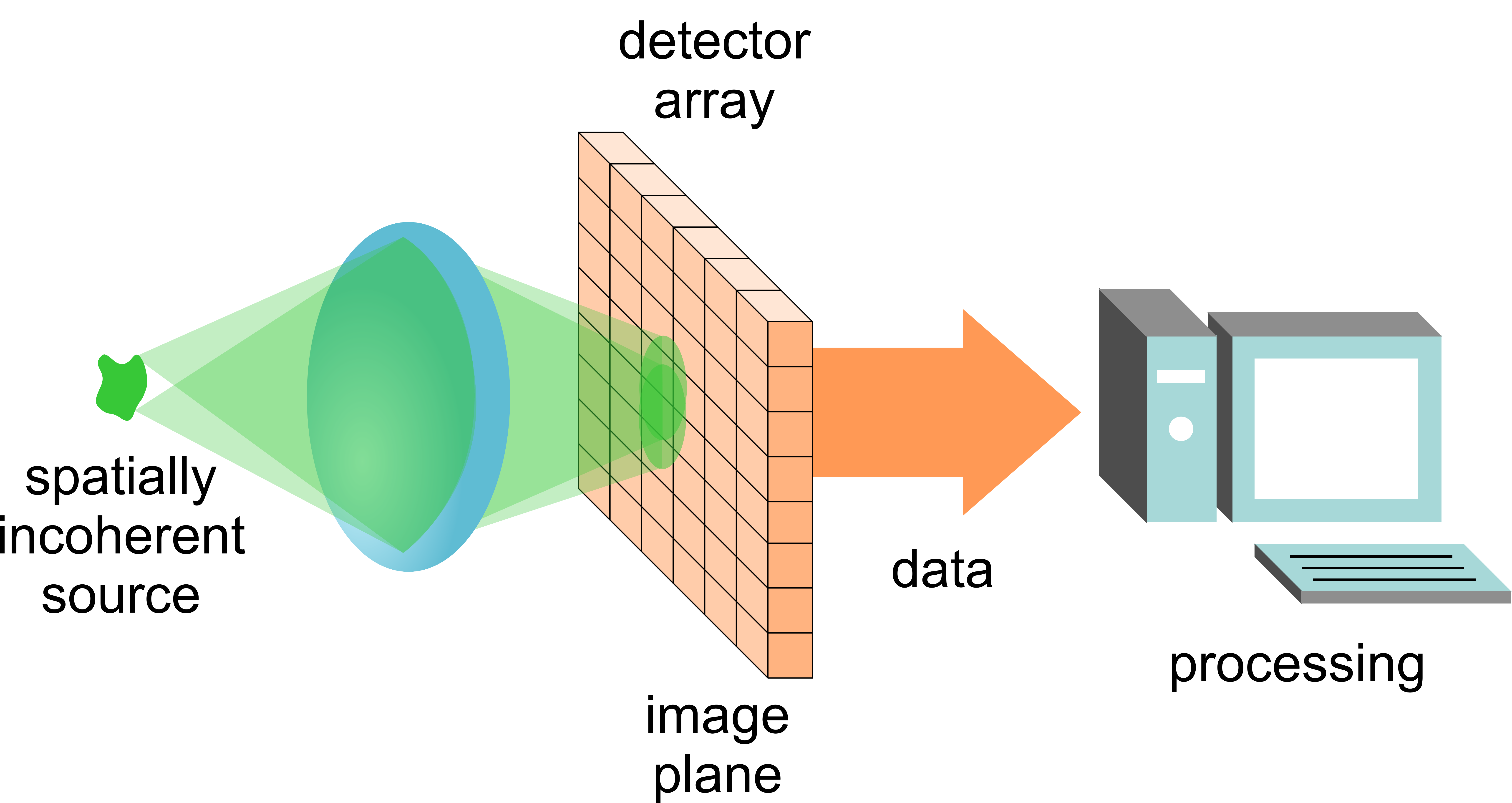}}
\caption{\label{direct_imaging}Basic setup of direct imaging.}
\end{figure}

The mathematics was formidable, but Helstrom managed to apply his
theory to a few simple scenarios of incoherent imaging. For example,
he studied the problem of locating an incoherent point source from
far-field measurements \cite{helstrom70}, but the result was
unsurprising: the quantum limit is close to the ideal performance of
direct imaging, which measures the intensity on the image plane, as
depicted by Fig.~\ref{direct_imaging}. A more intriguing problem he
studied was the decision between one or two incoherent sources
\cite{helstrom73b}. Helstrom computed the mathematical form of the
optimal measurement and the resulting error probabilities, but he did
not propose an experimental setup or show how much improvement the
optimal measurement could offer over existing imaging methods.
Helstrom himself was quite pessimistic \cite{helstrom73b}: ``The
optimum strategies required in order to attain the minimum error
probabilities calculated here require the measurement of certain
complicated quantum-mechanical projection operators, which, though
possible in principle, cannot be carried out by any known apparatus.''

Unfortunately, in all the problems studied by Helstrom, the
improvements predicted by his theory seemed modest at best, rendering
the question of quantum limits academic. Quantum opticians turned
their attention to nonclassical light sources
\cite{kolobov,dowling08,demkowicz15,taylor16,pirandola18,moreau19,fabre19},
while classical opticians turned their attention to near-field
microscopy \cite{betzig15,pendry04}, fluorescence control
\cite{betzig15,moerner15,hell15}, and computational imaging
\cite{villiers}. Helstrom's work on incoherent imaging was all but
forgotten.

Surprise came a few decades later. Applying quantum estimation theory
to the problem of resolving two incoherent point sources, we recently
discovered that substantial improvements via novel far-field
measurements are indeed possible \cite{tnl}. The theory has since been
generalized for an arbitrary number of sources
\cite{tsang17,tsang18a,dutton19,tsang19,zhou19,tsang19b,bonsma19}.
The implication is that, even for astronomy, where the sources are
inaccessible, the new techniques can enhance the resolution beyond the
limits of direct imaging---the de facto method developed by evolution
for eons and honed by opticians for centuries. I present in the
following an introduction to the breakthrough in Ref.~\cite{tnl}, as
well as the rapid theoretical
\cite{tsang17,tsang18a,dutton19,tsang19,zhou19,tsang19b,sliver,tnl2,nair_tsang16,lupo,tsang18,ant,lu18,rehacek17,yang17,kerviche17,chrostowski17,rehacek17a,rehacek18,backlund18,napoli19,yu18,prasad19,prasad19a,larson18,tsang_comment19,larson19,bonsma19,grace19,bisketzi19,lupo19,lee19,gefen19,hradil19,len20,lupo20}
and experimental
\cite{tang16,tham17,paur16,yang16,donohue18,parniak18,paur18,hassett18,zhou19a,paur19,wadood19,rehacek19}
advances that followed.

\section{Rayleigh's curse}
With two incoherent point sources, direct imaging, and photon shot
noise, many studies have shown that their separation becomes harder to
estimate if they violate Rayleigh's criterion
\cite{falconi67,tsai79,bettens,vanaert,ram}. The central tool used in
those studies is the Fisher information, which sets general lower
bounds called Cram\'er-Rao bounds on the parameter-estimation error
\cite{lehmann98}. The simplest Cram\'er-Rao bound ($\abbrev{CRB}$) is
\begin{align}
  \abbrev{MSE}(\theta) &\ge 
\abbrev{CRB}(\theta) \equiv \abbrev{FI}(\theta)^{-1},
\label{crb}
\end{align}
where $\abbrev{MSE}$ is the mean-square error of any unbiased
estimator, $\theta$ is the unknown parameter, and
$\abbrev{FI}(\theta)$ is the Fisher information; see
Appendix~\ref{sec_crb} for precise definitions. The error can reach
the Cram\'er-Rao bound in many situations, including an asymptotic
limit where the sample size approaches infinity, the noise can be
approximated as additive and Gaussian, and the maximum-likelihood
estimator is used \cite{lehmann98}. Thus, the Fisher information is a
useful measure of the sensitivity of the experiment to the unknown
parameter.

\begin{figure}[htbp!]
\centerline{\includegraphics[width=0.45\textwidth]{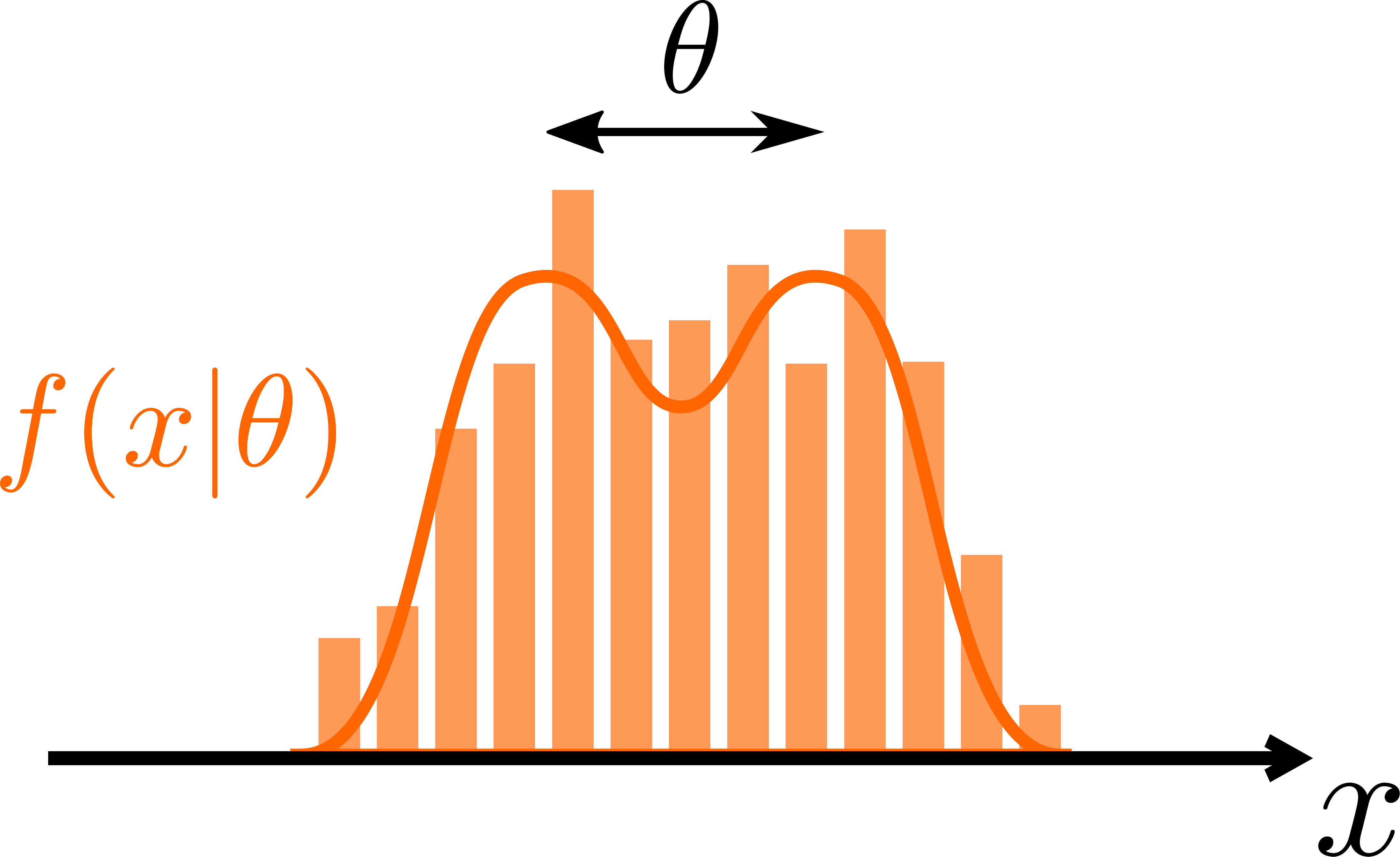}}
\caption{\label{psf_noisy3}The image of two point sources (histogram)
  is blurred by diffraction and corrupted by photon shot noise.
  $\theta$ denotes the separation between the sources, $f(x|\theta)$
  (solid curve) is the mean intensity, and $x$ is the image-plane
  coordinate.}
\end{figure}

Assume one-dimensional paraxial imaging \cite{goodman} for simplicity,
as illustrated by Fig.~\ref{psf_noisy3}, and Poisson noise, which is
an excellent approximation for both optical astronomy
\cite{feigelson,zmuidzinas03,goodman_stat} and fluorescence microscopy
\cite{pawley}.  The Fisher information becomes
\begin{align}
\abbrev{FI}^{({\rm direct})}(\theta) = C(\theta) N,
\label{fi_direct}
\end{align}
where $\theta$ here is the separation, $N$ is the average photon
number, and $C(\theta)$ is an $N$-independent prefactor that varies
with $\theta$.  $\theta$ and $C(\theta)$ are dimensionless if $\theta$
is normalized in Airy units (1 Airy unit is roughly
$\lambda/\abbrev{N.A.}$ where $\lambda$ is the wavelength and
$\abbrev{N.A.}$ is the numerical aperture, or $\lambda/D$ for angular
resolution, where $D$ is the aperture diameter \cite{pawley}).
Equation~(\ref{fi_direct}) was earlier suggested by many as a
fundamental measure of resolution for incoherent imaging
\cite{tsai79,bettens,vanaert,ram}.

\begin{figure}[htbp!]
\centerline{\includegraphics[width=0.45\textwidth]{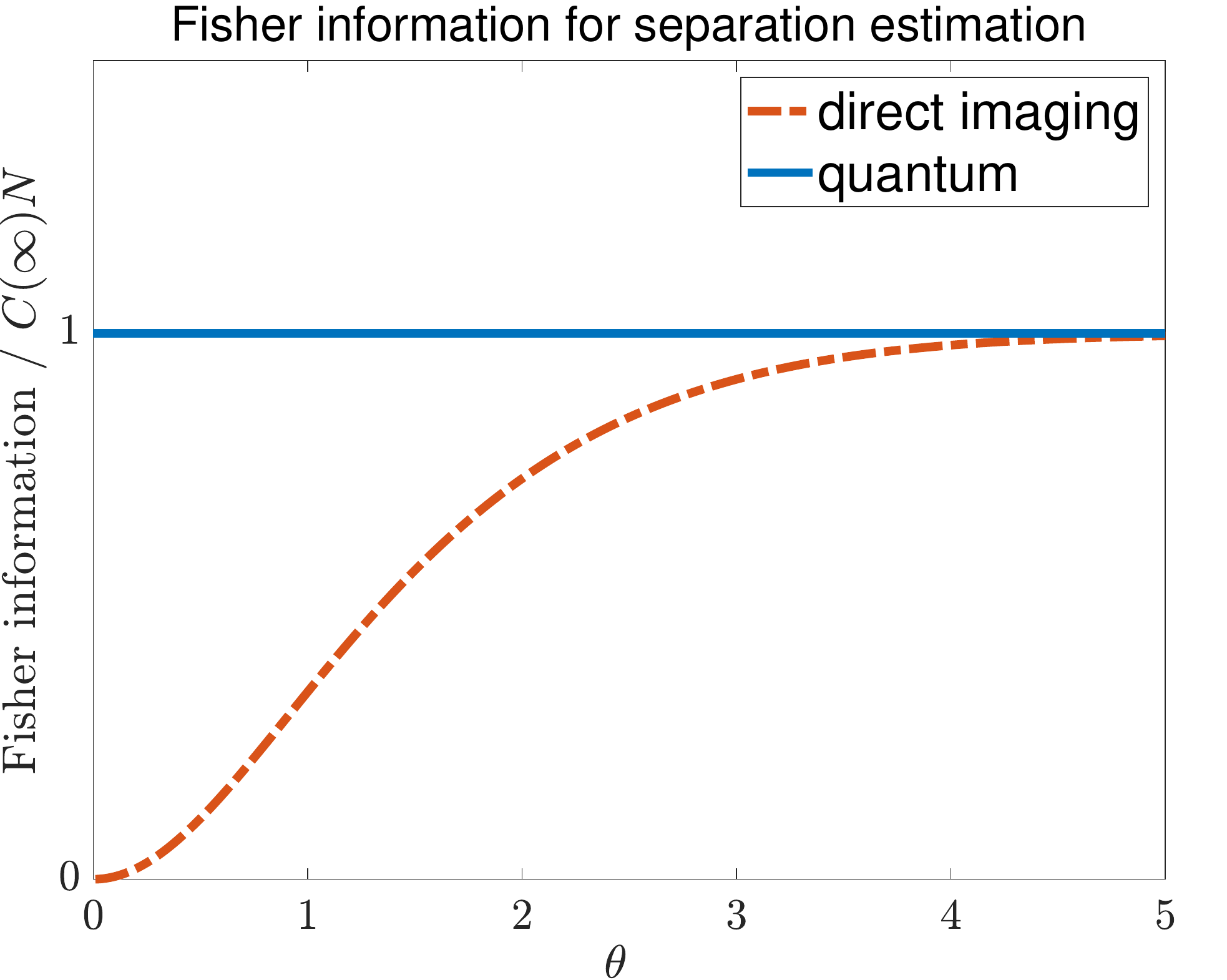}}
\caption{\label{fisher_separation}Fisher information for the
  estimation of the separation $\theta$ between two incoherent point
  sources, assuming a Gaussian point-spread function. With direct
  imaging, the information drops to zero for $\theta \to 0$, but the
  Helstrom information according to quantum estimation theory stays
  constant. }
\end{figure}

The details of $C(\theta)$ depend on the point-spread function, but
the general behavior is as follows: If the sources are well separated
relative to Rayleigh's criterion ($\theta \gg 1$), $C(\theta)$ is
relatively constant, but when $\theta$ is close to Rayleigh's
criterion or starts to violate it ($\theta \lesssim 1$), $C(\theta)$
decays to zero, causing the Cram\'er-Rao bound to blow up as
$\theta \to 0$.  In other words, there is a progressive penalty on the
Fisher information for the violation of Rayleigh's criterion, as
illustrated by Fig.~\ref{fisher_separation} for a Gaussian
point-spread function. In Ref.~\cite{tnl}, we called this penalty
Rayleigh's curse to distinguish it from Rayleigh's
criterion---sub-Rayleigh sources are resolvable, but the more they
violate Rayleigh's criterion, the harder it gets to estimate their
separation.

\section{Dispelling Rayleigh's curse}
Rayleigh's curse happens if we measure the intensity on the image
plane, but what if we allow any quantum measurement that may be
sensitive to the phase as well?  To find the quantum limit, we can use
a quantum version of the Fisher information proposed by Helstrom
\cite{helstrom}, which sets an upper bound on the Fisher information
for any measurement \cite{nagaoka89,braunstein}, as elaborated in
Appendix~\ref{sec_hi}.  We found that the Helstrom information
($\abbrev{HI}$) for the separation estimation problem is given by
\cite{tnl}
\begin{align}
\abbrev{FI}(\theta) \le \abbrev{HI}(\theta) = C(\infty)N.
\label{hi}
\end{align}
Remarkably, $\abbrev{HI}(\theta)$ is constant regardless of the
separation and completely free of Rayleigh's curse, as plotted in
Fig.~\ref{fisher_separation}.

The constant Helstrom information would be no surprise if it were
simply a loose upper bound; the million-dollar question is whether one
can find a measurement that attains the limit. Mathematical studies
following Helstrom's work have shown in general that a quantum-limited
measurement should exist, at least in the limit of infinite sample
size \cite{hayashi05,fujiwara2006}.  The mathematics offers little
clue to the experimental implementation, however, and finding one in
quantum estimation theory is often a matter of educated guessing.

\begin{figure}[htbp!]
\centerline{\includegraphics[width=0.45\textwidth]{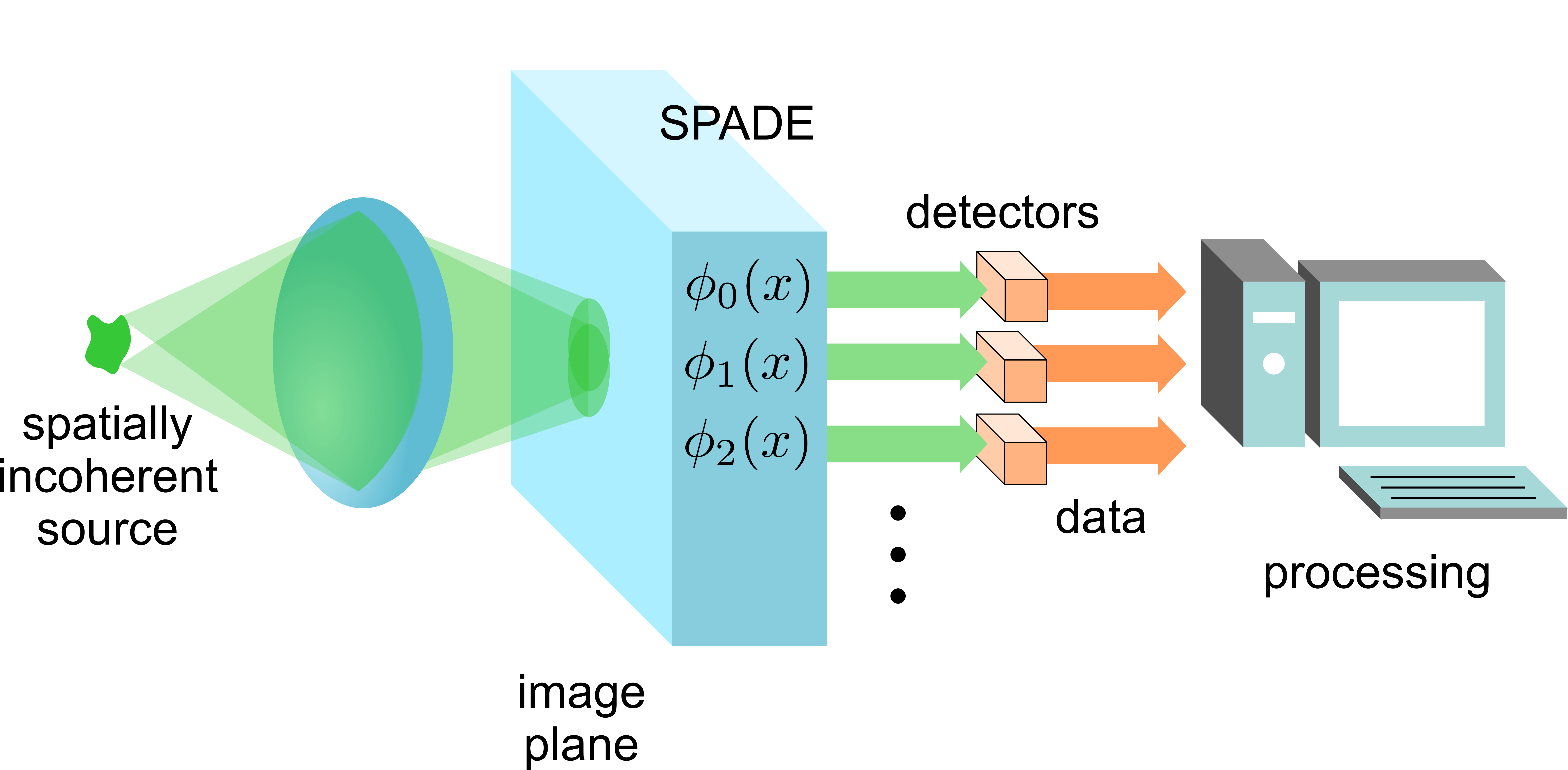}}
\caption{\label{spade}Basic setup of SPADE for incoherent imaging.}
\end{figure}

Luckily we found one. Assuming a Gaussian point-spread function, we
found that sorting the light on the image plane in terms of the
Hermite-Gaussian modes, followed by photon counting in each mode, can
lead to a Fisher information given by \cite{tnl}
\begin{align}
  \abbrev{FI}^{({\rm SPADE})}(\theta) &= C(\infty)N,
\label{fi_spade}
\end{align}
which attains the quantum limit and is free of Rayleigh's curse for
all $\theta$. Figure~\ref{spade} illustrates the setup.  We called the
measurement spatial-mode demultiplexing with the acronym SPADE, to
follow the convention of giving catchy acronyms to superresolution
methods \cite{moerner15}.  Numerical simulations have shown that SPADE
combined with a judicious estimator can give an error very close to
the quantum bound $1/\abbrev{HI}$ and substantially lower than that
achievable by direct imaging \cite{tnl,tsang18}. Further studies have
proposed measurements that work for other point-spread functions
\cite{tnl,sliver,rehacek17,kerviche17}.

\section{How SPADE works}
To understand how SPADE can beat direct imaging and achieve the
quantum limit, it is helpful to consider a simplified model of thermal
light \cite{tnl} that is valid for optical frequencies and beyond, as
described in the following. The model may sound heuristic, but it is
possible to derive it from a quantum formalism by assuming a thermal
quantum state \cite{mandel}, the paraxial optics model
\cite{yuen_shapiro1}, and an ``ultraviolet'' limit, as elaborated in
Appendix~\ref{sec_thermal}.

Treat each photon on the image plane as a quantum particle with
wavefunction $\psi(x)$, where $x$ is the image-plane coordinate
normalized with respect to the magnification factor \cite{goodman}.
Direct imaging corresponds to a measurement of its position, obeying
the probability density
\begin{align}
f(x) = |\psi(x)|^2,
\end{align}
by virtue of Born's rule. It is also possible to measure the particle
in any other orthonormal basis $\{\phi_q(x): q \in \mathbb N_0\}$, and
the probability of finding the photon in the $q$th spatial mode is
\begin{align}
g_q &= \abs{\intall dx \phi_q^*(x)\psi(x)}^2.
\end{align}
For incoherent imaging, the wavefunction of each photon is
$\psi(x-X)$, where $\psi$ is determined by the point-spread function
of a diffraction-limited imaging system and the displacement $X$
depends on the position of the point source that emits the photon.
Denoting the density of the incoherent sources as $F(X)$, $X$ can be
regarded as a random variable with $F(X)$ as its probability density.
For direct imaging, the probability density on the image plane becomes
\begin{align}
f(x) &= \intall dX |\psi(x-X)|^2 F(X),
\label{direct}
\end{align}
which agrees with the classical theory of incoherent imaging
\cite{goodman}. In general, the probability of finding the photon in
the $\phi_q(x)$ mode is
\begin{align}
g_q &= \intall dX \abs{\intall dx \phi_q^*(x)\psi(x-X)}^2 F(X).
\label{Pq}
\end{align}
If we treat the arrivals of the photons at the spatial modes as a
temporal Poisson process, then the photon counts integrated over time
are independent Poisson random variables, each with mean and variance
given by $N g_q$, where $N$ is the average photon number in all modes.
For direct imaging, the photon statistics should be treated as a
spatial Poisson process with mean intensity $Nf(x)$
\cite{snyder_miller}.

\begin{figure}[htbp!]
\centerline{\includegraphics[width=0.45\textwidth]{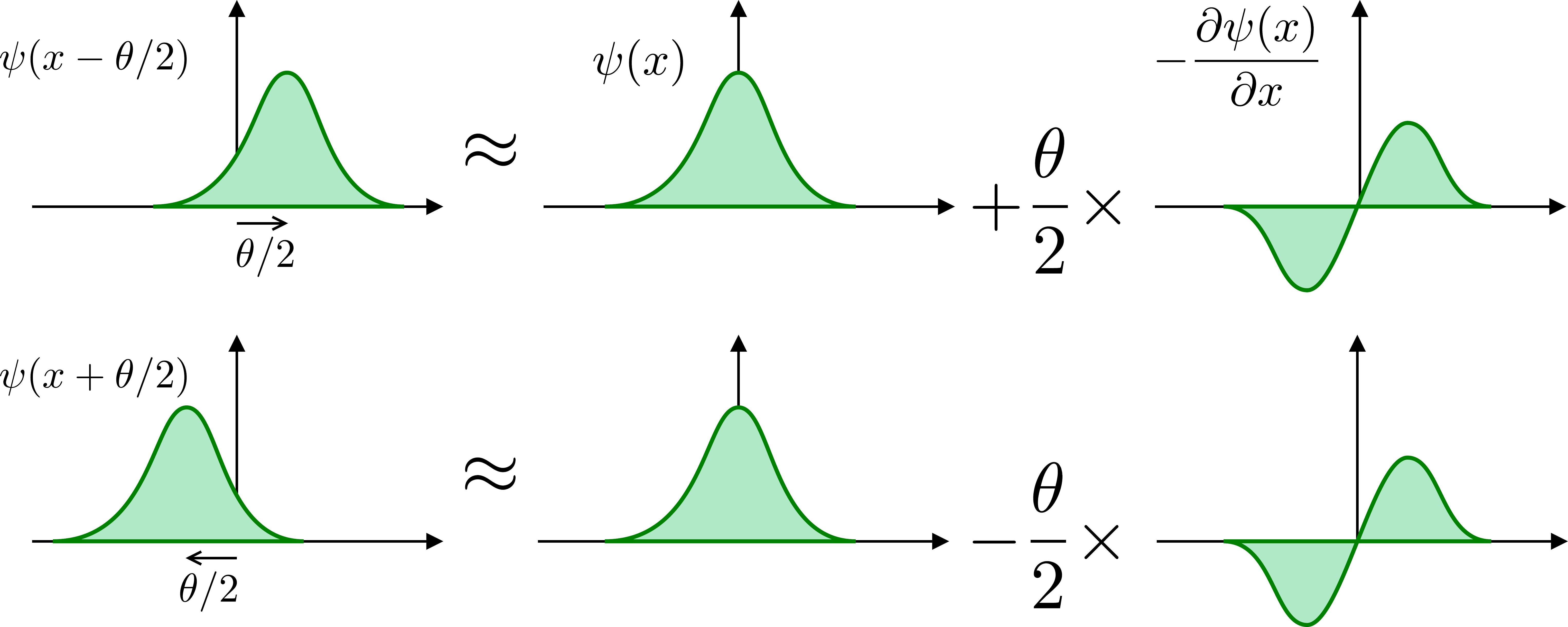}}
\caption{\label{spade_explain1}The wavefunction due to each point
  source can be decomposed in terms of the fundamental mode $\psi(x)$
  and the derivative mode $-\partial\psi(x)/\partial x$ for
  $\theta\ll 1$. For incoherent point sources, the total energy in the
  derivative mode consists of the incoherent contributions from the
  sources ($\propto (\theta/2)^2+(-\theta/2)^2 = \theta^2/2$).  Thus
  the derivative mode contains the signal about $\theta$, while the
  fundamental mode acts as a background noise.}
\end{figure}

Consider two point sources, one at $X = -\theta/2$ and one at
$X = \theta/2$ such that
$F(X) = [\delta(X-\theta/2)+\delta(X+\theta/2)]/2$. If their
separation is deeply sub-Rayleigh ($\theta \ll 1$), the wavefunctions
can be approximated as
\begin{align}
\psi\bk{x\pm\frac{\theta}{2}} &\approx \psi(x) \pm\frac{\theta}{2}
\parti{\psi(x)}{x},
\label{taylor}
\end{align}
as depicted by Fig.~\ref{spade_explain1}. If $\psi(x)$ is even,
$\partial \psi(x)/\partial x$ is odd, and they can be regarded as two
orthogonal modes. To the first order, the mean photon count in the
fundamental $\psi(x)$ mode is insensitive to the parameter $\theta$,
while the mean count in the derivative mode is the incoherent sum of
the contributions from the two sources, or
$\propto (\theta/2)^2 + (-\theta/2)^2 = \theta^2/2$.  If the sources
were coherent and in-phase instead, their contributions to the
derivative mode would cancel each other, leading to a much reduced
signal \cite{localization}.  In other words, the incoherence plays a
key role in retaining a significant signal in the first order, and
SPADE can extract this signal by measuring the derivative mode.

Another reason that SPADE can outperform direct imaging has to do with
the fundamental mode $\psi(x)$. It contains little signal, but it
overlaps spatially with the derivative mode and contributes a
background to the spatial intensity measured by direct imaging,
increasing the variances of the photon counts at each pixel. By
projecting the fundamental mode into a different channel, SPADE
filters out this background noise and substantially improves the
signal-to-noise ratio.

The heuristic discussion so far can be made more rigorous by
considering the Fisher information and the Cram\'er-Rao bounds.
Assume that the object distribution $F(X|\theta)$ and therefore
$f(x|\theta)$ and $g_q(\theta)$ depend on $\theta$. For the spatial
Poisson process from direct imaging, the Fisher information is
\cite{snyder_miller}
\begin{align}
\abbrev{FI}^{({\rm direct})}(\theta) &= 
N \intall dx \frac{1}{f(x|\theta)}\Bk{\parti{f(x|\theta)}{\theta}}^2.
\label{fi_direct2}
\end{align}
For separation estimation with $\theta \ll 1$,
\begin{align}
f(x|\theta) \approx |\psi(x)|^2+\frac{\theta^2}{8}\partit{|\psi(x)|^2}{x}.
\label{fapprox}
\end{align}
The denominator in Eq.~(\ref{fi_direct2})
approaches $|\psi(x)|^2$ as $\theta \to 0$, meaning that the
fundamental mode is the major noise contributor, and the Fisher
information approaches zero as $\theta \to 0$. For discrete Poisson
variables on the other hand, the Fisher information is
\begin{align}
\abbrev{FI}(\theta) &= N 
\sum_q \frac{1}{g_q(\theta)}\Bk{\parti{g_q(\theta)}{\theta}}^2.
\label{fi}
\end{align}
For separation estimation, as long as $\phi_1(x)$ is orthogonal to
$\psi(x)$ and has significant overlap with the derivative mode,
$g_1(\theta) \propto \theta^2$ for $\theta \ll 1$, leading to a
nonzero $[\partial g_1(\theta)/\partial \theta]^2/g_1(\theta)$ as
$\theta \to 0$.

To summarize, SPADE relies on the subtle interplay between the
coherence induced by diffraction, the incoherence of the sources, and
the signal-dependent nature of photon shot noise. It would have been
difficult to discover such a fortuitous possibility via conventional
wisdom alone, but quantum estimation theory---and quantum information
theory in general---have the advantage of being oblivious to
conventional wisdom. The mathematics may look daunting, but it can
sometimes give rise to new physics beyond our imagination.

\section{Implementations of SPADE}
To implement SPADE, different spatial modes should be coupled into
physically separate channels before detection. This in principle
requires only linear optics \cite{morizur}, but the most efficient
implementation remains unclear.  Many methods have been proposed and
demonstrated, particularly for the purpose of mode-division
multiplexing in optical communication \cite{fabre19}. Here I highlight
a few methods that have been experimentally demonstrated for the
two-point resolution problem.

\subsection{Interferometry}
Nair proposed an interferometer called SLIVER (superlocalization via
image-inversion interferometry) that can in principle achieve a
quantum-limited Fisher information for $\theta \to 0$ and any even
point-spread function \cite{sliver}.  Although image-inversion
interferometry has earlier been proposed and demonstrated to combat
atmospheric turbulence for astronomy \cite{roddier88} and to achieve a
modest resolution improvement for general confocal microscopy
\cite{wicker07,wicker09,weigel,weigel11}, its extraordinary precision
for sub-Rayleigh resolution was hitherto not recognized.

\begin{figure}[htbp!]
\centerline{\includegraphics[width=0.45\textwidth]{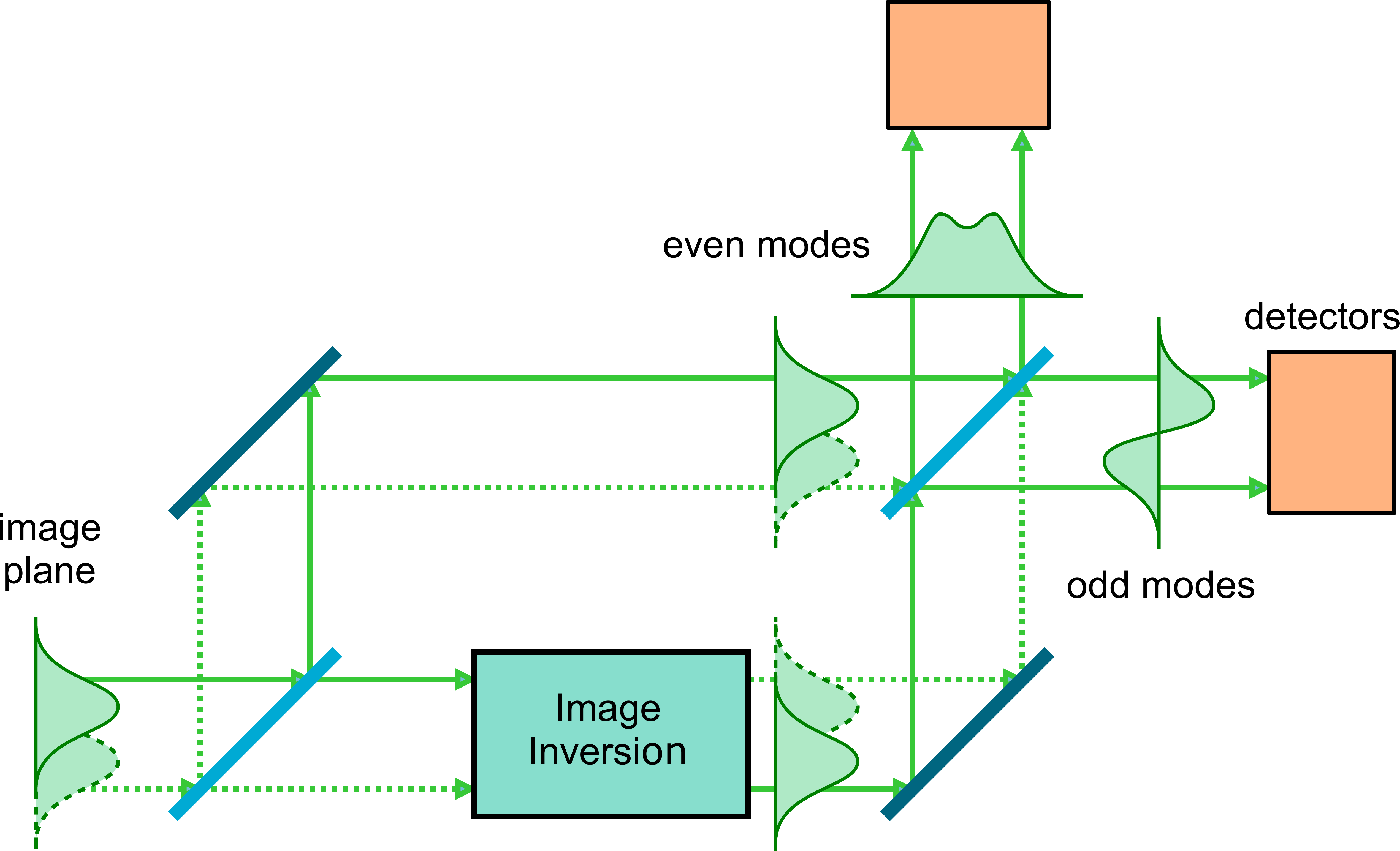}}
\caption{\label{sliver}An image-inversion interferometer.  Through the
  inversion and the interference, the even modes are coupled to one
  port while the odd modes are coupled to the other port.}
\end{figure}

The setup, depicted by Fig.~\ref{sliver}, consists of a two-arm
interferometer with spatial inversion in one arm. The inversion can be
implemented via mirrors, lenses, or a Dove prism for example. As a
result of the inversion and the interference at the second
beamsplitter, all the even modes on the image plane are routed to one
output port while the odd modes are routed to the other port. Hence,
the fundamental mode $\psi(x)$, as long as it is even, is separated
from the odd derivative mode, which is detected at the other
port. Tang, Durak, and Ling reported a proof-of-concept demonstration
of SLIVER \cite{tang16}, although their reported errors were not close
to the quantum limit. Larson and coworkers recently reported a
common-path configuration of the interferometer that may be more
stable \cite{larson19a}.

SLIVER works best for sub-Rayleigh separations but is suboptimal for
larger separations. A variant of SLIVER called pix-SLIVER replaces the
detectors by detector arrays and can work better for larger
separations \cite{nair_tsang16}. Another way to generalize SLIVER is
to think of image inversion as a special case of fractional Fourier
transform (FRFT). A tree of FRFT interferometers, with the
image-inversion interferometer at its root, can sort the
Hermite-Gaussian modes and implement SPADE \cite{xue01}. The
interferometer-tree concept can be generalized to sort in any other
basis if appropriate mode-dependent phases can be introduced
\cite{abouraddy12,martin17}.

Along this direction, Hassett and coworkers demonstrated a Michelson
interferometer with variable FRFT in one arm and used it to infer the
Hermite-Gaussian-mode spectrum $g_q$ of a shifted Gaussian beam
\cite{hassett18}. They suggested that the setup could be useful for
estimating sub-Rayleigh separations, although its statistical
performance remains to be studied.  In another work, Zhou and
coworkers demonstrated a binary radial-mode sorter that is also based
on FRFT interferometry and used it to enhance the estimation of the
axial separation between two sources \cite{zhou19a}.

\subsection{SPLICE}
\begin{figure}[htbp!]
\centerline{\includegraphics[width=0.45\textwidth]{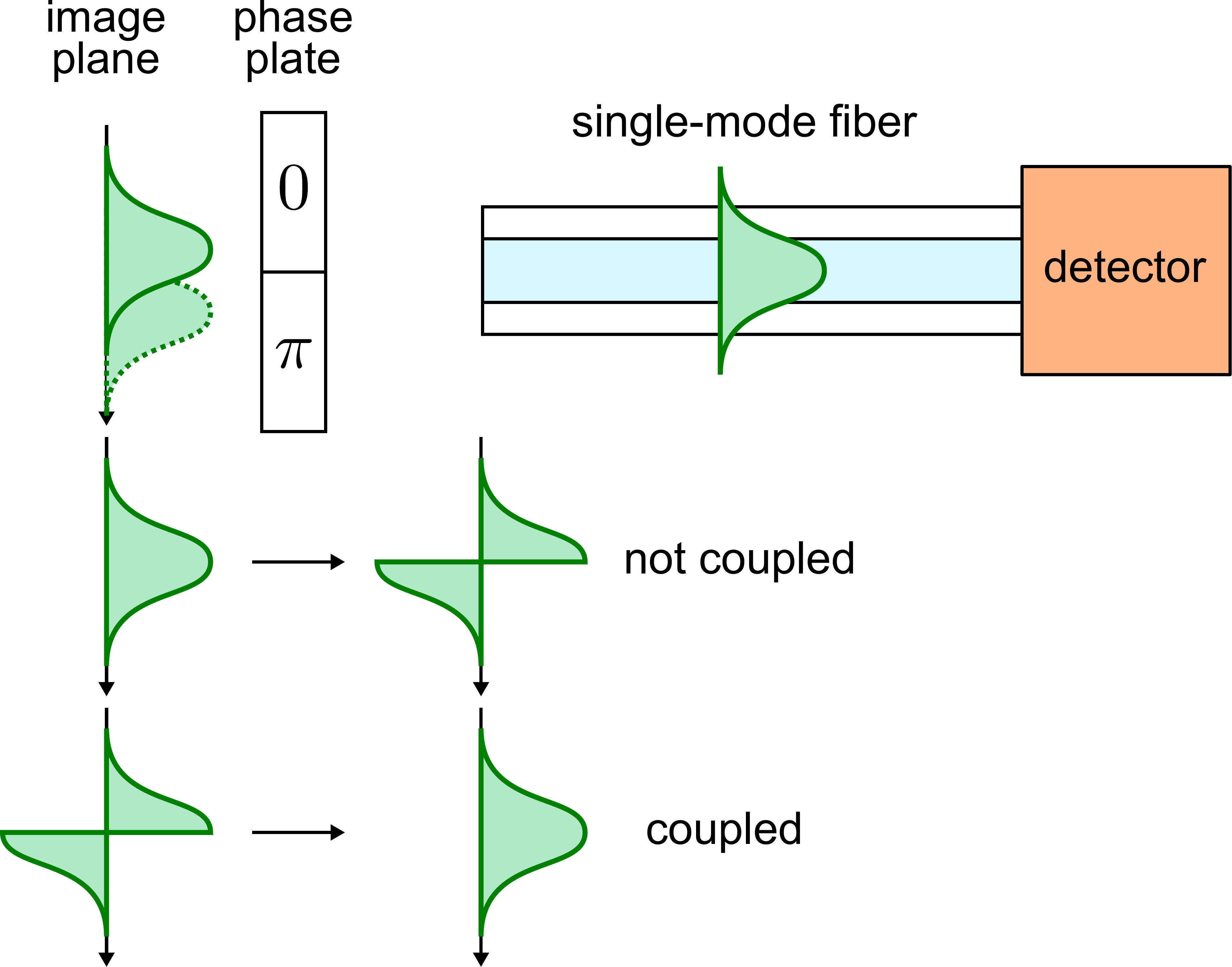}}
\caption{\label{splice}Setup and principle of SPLICE
  \cite{tham17}. The phase plate introduces a $\pi$ phase shift to
  half of the image plane relative to the other half. Only the odd
  mode that has been converted by the phase plate to the fiber mode is
  coupled into the fiber and detected.}
\end{figure}

Tham, Ferretti, and Steinberg proposed an elegant setup called SPLICE
(super-resolved position localization by inversion of coherence along
an edge) to capture the derivative mode \cite{tham17}. SPLICE consists
of a phase plate that introduces a $\pi$ phase shift to half of the
image plane and a single-mode fiber, as illustrated by
Fig.~\ref{splice}. An odd mode on the image plane is thus coupled into
the fiber and detected, while all other modes orthogonal to it are
rejected by the fiber. Despite the imperfect match between the odd
mode and the derivative mode, Tham and coworkers were still able to
demonstrate a mean-square error around five times the quantum bound
and a significant improvement over direct imaging \cite{tham17}.

The use of phase plates is, of course, routine in phase-contrast
microscopy \cite{goodman,lohmann98}, while the use of a half-plane
$\pi$-phase plate specifically also has a long history in coherent
imaging \cite{wolter50,lohmann98}. The important distinctions here are
that we are dealing with incoherent sources, the phase plate is placed
at the image plane, and there is a fiber that performs judicious
spatial-mode selection.

\subsection{Holograms}
A hologram is capable of performing a spatial matched filter, and it
can be designed such that the diffracted intensities at specific
points in the far field are proportional to the modal spectrum $g_q$
\cite{goodman,forbes16}. The use of such a hologram for separation
estimation was demonstrated by Pa\'ur and coworkers
\cite{paur16}. Their reported mean-square errors were around twice the
quantum bound, but it is important to note that they scaled the
quantum bound with respect to the diffracted photon number, not the
photon number before the hologram, meaning that the result did not
take into account the low diffraction efficiency of their hologram.
Efficient SPADE is possible with multiple holograms, however
\cite{fabre19}.

\subsection{Point-spread-function shaping}
In the context of direct imaging, the approximation given by
Eq.~(\ref{fapprox}) for $\theta \ll 1$ leads to
\begin{align}
\abbrev{FI}^{({\rm direct})} &\approx 
\frac{N\theta^2}{16} \intall dx 
\frac{[\partial^2|\psi(x)|^2/\partial x^2]^2}
{|\psi(x)|^2 + (\theta^2/8)\partial^2|\psi(x)|^2/\partial x^2}.
\label{fi_approx2}
\end{align}
It is often assumed \cite{bettens,vanaert} that this
can be approximated by
\begin{align}
\abbrev{FI}^{({\rm direct})} &\approx 
\frac{N\theta^2}{16} \intall dx \frac{1}{|\psi(x)|^2}
\Bk{\partit{|\psi(x)|^2}{x}}^2,
\label{fi_approx3}
\end{align}
which scales quadratically with $\theta$.  This is indeed true if
$|\psi(x)|^2$ is Gaussian, but it turns out that the integral in
Eq.~(\ref{fi_approx3}) may not converge if $|\psi(x)|^2$ has zeros,
and one must go back to Eq.~(\ref{fi_approx2}), which can give a
linear scaling of $\abbrev{FI}^{({\rm direct})}$ with $\theta$
instead. Pa\'ur and coworkers exploited this phenomenon by
introducing a signum phase mask at the pupil plane of a direct-imaging
system, changing $\psi(x)$ from a Gaussian to an odd function with a
zero in the middle \cite{paur18}. Although the resulting Fisher
information still approaches zero for $\theta \to 0$, they were able
to demonstrate a significant improvement of the estimation accuracy
with a simple change. Further experiments along the same line for
spectroscopy have recently been reported \cite{paur19}.

\subsection{Heterodyne}
Given the experimental difficulties of performing efficient SPADE, a
seemingly appealing alternative is to perform heterodyne detection of
the derivative mode by interfering the light with a shaped reference
beam on a detector, as demonstrated by Yang and coworkers
\cite{yang16}. It was later found, however, that the homodyne or
heterodyne Fisher information still suffers from Rayleigh's curse for
weak thermal light \cite{yang17}. This can be attributed to the
constant vacuum noise that plagues a heterodyne or homodyne detection
regardless of the signal, compared with the Poisson variance that
reduces with the signal for photon counting. A similar problem was
discovered earlier in the context of stellar interferometry
\cite{townes,stellar}. The surprisingly poor performance of heterodyne
detection demonstrates the importance of analyzing a measurement using
rigorous quantum optics as well as statistics, even when dealing with
classical light, to ensure an acceptable statistical performance.

\subsection{Sum-frequency generation}
Donohue and coworkers implemented SPADE in the time or frequency
domain for estimating the separation between optical pulses via an
interesting nonlinear-optical technique: sum-frequency generation
\cite{donohue18}. If the light is combined with a strong
local-oscillator pulse in a second-order nonlinear medium with the
right phase matching, the Hamiltonian of the sum-frequency generation
is the same as that of linear optics \cite{eckstein11}, and a temporal
or spectral mode projection can be implemented if the local oscillator
has the desired mode shape and the up-converted signal is measured.
While the efficiency of their measurement was only 0.7\%, the
principle was clearly demonstrated in their experiment.

\subsection{Two-photon measurement}
Last but not the least, I should mention an even more radical proposal
by Parniak and coworkers, which uses a two-photon measurement to
estimate the centroid and the separation of two sources simultaneously
near the quantum limit \cite{parniak18}. Its applicability to usual
light sources is questionable, but it demonstrates the fact that our
model of linear optics and Poisson statistics does not encompass all
the possibilities offered by quantum mechanics, and there exist
multiphoton measurements that can offer advantages in multiparameter
estimation, at least in principle.

\section{Extended sources}

\subsection{Estimation of the second moment}
\begin{figure}[htbp!]
\centerline{\includegraphics[width=0.45\textwidth]{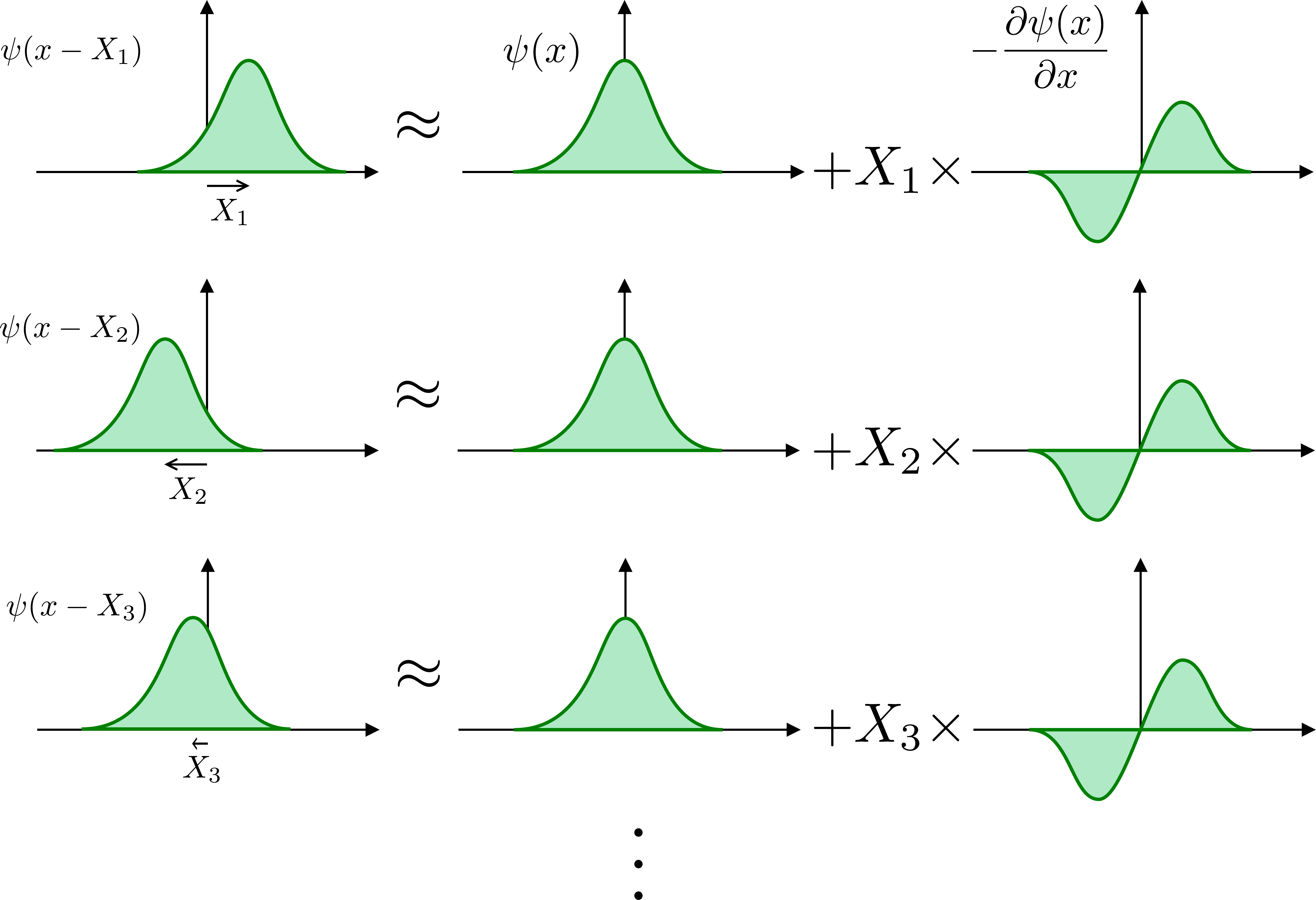}}
\caption{\label{spade_explain_multi}The wavefunction due to each point
  source can be decomposed in terms of the fundamental mode $\psi(x)$
  and the derivative mode $-\partial\psi(x)/\partial x$ for
  $|X| \ll 1$. For multiple incoherent point sources, the total
  energy in the derivative mode consists of the incoherent
  contributions from the sources ($\propto X_1^2+X_2^2+X_3^2+\dots$
  for equally bright sources). In other words, the energy is
  proportional to the second moment of the source distribution.}
\end{figure}

While the two-point problem is historic and significant, it has rather
limited applications, and the important next step is to apply the
concepts developed so far to more general objects. Suppose now that
the number of point sources is arbitrary, and the object intensity is
given in general by $F(X)$.  Similar to the sub-Rayleigh approximation
earlier, here I focus on a subdiffraction regime where the object
width around $X = 0$, defined as $\Delta$, is much smaller than the
width of the point-spread function, or $\Delta \ll 1$. Otherwise,
$F(X)$ is assumed to be unknown to the experimenter.  Similar to
Eq.~(\ref{taylor}), the photon wavefunction due to each point $X$
within the object can be approximated as
\begin{align}
\psi(x-X) &\approx \psi(x) - X\parti{\psi(x)}{x}.
\label{taylorX}
\end{align}
Summing the incoherent contributions from all the points via
Eq.~(\ref{Pq}), the mean photon count in the derivative mode
$\phi_1(x)\propto \partial\psi(x)/\partial x$ is
\begin{align}
N g_1 &\approx N c_1^2  \intall dX X^2 F(X),
\end{align}
where $c_1$ is a constant and $\intall dX X^2F(X)$ is the second
moment of $F(X)$. Figure~\ref{spade_explain_multi} illustrates this
concept for multiple point sources.  Thus we can expect SPADE to
enhance the estimation of the second moment for any subdiffraction
object in the same way it enhances the two-point resolution.  As the
second moment can be related to the width of $F(X)$, it should not be
surprising that SPADE can also enhance the estimation of the object
size \cite{tsang17,dutton19}.

\subsection{Even moments}
To go another step further, let us expand $\psi(x-X)$ up to the $q$th
order. It is more convenient to work in the spatial frequency domain,
as defined by
\begin{align}
  \psi(x) &\to \Psi(k) =
\frac{1}{\sqrt{2\pi}}\intall dx \psi(x)\exp(-ikx),
\label{Psi}
  \end{align}
which leads to
\begin{align}
  \psi(x-X) &\to \exp(-ikX)\Psi(k) \approx 
              \sum_{p=0}^q \frac{(-ikX)^p}{p!}\Psi(k).
\label{taylor_k}
\end{align}
A natural orthonormal basis that includes the fundamental mode
$\psi(x)\to\Psi(k)$ and the derivative mode
$-\partial\psi(x)/\partial x \to -ik\Psi(k)$ can be defined as
\cite{rehacek17}
\begin{align}
\BK{\phi_q(x) \to \Phi_q(k) = (-i)^qb_q(k)\Psi(k): q \in \mathbb N_0},
\label{pad}
\end{align}
where $\{b_q(k)\}$ are the orthogonal polynomials obtained by applying
the Gram-Schmidt process \cite{debnath05} to monomials
$\{1,k,k^2,\dots\}$ with respect to the weighted inner product
\cite{dunkl}
\begin{align}
\avg{u(k),v(k)} \equiv \intall dk |\Psi(k)|^2u^*(k)v(k),
\label{wip}
\end{align}
leading to
$\avg{b_q(k),b_p(k)} = \intall dk \Phi_q^*(k)\Phi_p(k) = \delta_{qp}$.
Appendix~\ref{sec_gs} gives a brief review of the Gram-Schmidt
process.  The basis $\{\phi_q(x)\}$ is called the
point-spread-function-adapted basis \cite{rehacek17}, or the PAD basis
for short \cite{tsang18a}. For example, if $|\Psi(k)|^2$ is Gaussian,
then $\{b_q(k)\}$ are the Hermite polynomials.  An important property
of $b_q(k)$ that follows from the Gram-Schmidt process is that
$\avg{b_q(k),k^p} = 0$ if $p < q$. The overlap function in
Eq.~(\ref{Pq}) becomes
\begin{align}
&\quad \intall dx \phi_q^*(x)\psi(x-X) 
\nonumber\\
&\approx 
\sum_{p=0}^q \frac{(-iX)^p}{p!}\intall dk \Phi_q^*(k)\Psi(k) k^p
\label{overlap1}\\
&=
\sum_{p=0}^q \frac{(-iX)^p}{p!}i^q \Avg{b_q(k),k^p}
= c_q X^q,
\label{overlap2}
\end{align}
where $c_q$ is a real constant. In other words, $\Phi_q(k)$ is orthogonal
to all the terms in Eq.~(\ref{taylor_k}) except the last $q$th-order
term (and the neglected higher-order terms). The mean photon count
given by Eq.~(\ref{Pq}) becomes
\begin{align}
N g_q &\approx N c_q^2\intall dX X^{2q} F(X).
\end{align}
Similar to the relation between the derivative mode and the second
moment, each PAD mode can access an even moment while rejecting the
background noise from all the lower moments \cite{tsang18a}. Hence,
SPADE with respect to the PAD basis can be expected to enhance the
estimation of all even moments.

If $\psi(x)$ is Gaussian, the PAD basis becomes the Hermite-Gaussian
basis, and its sensitivity to even moments was noted in
Refs.~\cite{yang16,tsang17}.  The general PAD basis was proposed in
Refs.~\cite{rehacek17,kerviche17} for the two-point problem and
applied to general imaging in Refs.~\cite{tsang18a,zhou19,tsang19b}.
The use of SPLICE for moment estimation was recently proposed by
Bonsma-Fisher and coworkers \cite{bonsma19}.

\subsection{Error analysis}
Define the moment parameters as
\begin{align}
\theta_\mu &= \intall dX X^\mu F(X),
\label{moment}
\end{align}
where $\mu \in \mathbb N$ denotes the moment order.
Appendix~\ref{sec_multi} introduces the multiparameter-estimation
theory in more detail. The mean and variance of the photon count $n_q$
in each PAD mode is
\begin{align}
Ng_q &\approx N c_q^2\theta_{2q},
\end{align}
so the estimator $\check\theta_{2q} = n_q/(Nc_q^2)$ is approximately
unbiased, and the mean-square error is
\cite{tsang17,tsang18a,tsang19b}
\begin{align}
  \abbrev{MSE}_{2q}^{({\rm SPADE})} 
&\approx \frac{\theta_{2q}}{Nc_q^2} = \frac{O(\Delta^{2q})}{N},
\label{mse_even}
\end{align}
where the subscript $2q$ denotes the error for the $\theta_{2q}$
parameter, the big-O notation denotes terms on the order of the
argument, and $\theta_\mu = O(\Delta^\mu)$. For direct imaging on the
other hand, the Cram\'er-Rao bound for any moment is
\cite{tsang17,tsang18a,tsang19b}
\begin{align}
\abbrev{MSE}_\mu^{({\rm direct})} &\ge
\abbrev{CRB}_{\mu}^{({\rm direct})} = \frac{O(1)}{N},
\label{CRBdirect}
\end{align}
so SPADE can achieve much lower errors for the even moments in the
$\Delta \ll 1$ subdiffraction regime. The exact Cram\'er-Rao bounds
for both SPADE and direct imaging, as well as the unbiased estimators
to achieve them, have been derived recently in Ref.~\cite{tsang19b}
via semiparametric methods and are consistent with the approximate
results here.

As large as the enhancement seems, the signal-to-noise ratio
($\abbrev{SNR}$), defined as
\begin{align}
\abbrev{SNR}_\mu  &\equiv \frac{\theta_\mu^2}{\abbrev{MSE}_\mu},
\end{align}
offers a more sobering perspective, as the signal
$\theta_\mu^2 = O(\Delta^{2\mu})$ is an even smaller number. For SPADE
and even moments, the SNR turns out to be equal to the mean photon
count in a PAD mode, or
\begin{align}
\abbrev{SNR}_{2q}^{({\rm SPADE})} &\approx Ng_q = N O(\Delta^{2q}),
\label{snr_even}
\end{align}
which decreases for smaller $\Delta$ and higher moments. The
degradation of the SNR can be attributed to the inherently low
efficiency of a subdiffraction source coupling into a higher-order
mode. While this shows that SPADE has its own limitations, the fact
remains that direct imaging is even worse, with a SNR given by
\begin{align}
\abbrev{SNR}_{\mu}^{({\rm direct})} = N O(\Delta^{2\mu}),
\end{align}
which is $N O(\Delta^{4q})$ for $\mu = 2q$. With enough photons, the
enhancements offered by SPADE can still be useful, especially for the
lower moments.

\subsection{Odd moments}
To estimate an odd moment, consider projections into the pair of
so-called iPAD modes
\begin{align}
\phi_q^{(\pm)}(x) &= \frac{\phi_q(x)\pm \phi_{q+1}(x)}{\sqrt{2}},
\end{align}
which result from the interference of two adjacent PAD modes
\cite{tsang18a}. It makes intuitive sense that, if each
$\phi_q$ mode is sensitive to the $2q$th moment, then a superposition
of two adjacent PAD modes should be sensitive to an odd moment
in-between.  Expanding $\psi(x-X)$ up to the $(q+1)$th order and
following the same steps as Eqs.~(\ref{overlap1}) and
(\ref{overlap2}), the overlap function becomes
\begin{align}
\int dx \phi_q^{(\pm)*}(x)\psi(x-X) &\approx 
\frac{1}{\sqrt{2}} \bk{c_q X^q\pm c_{q+1}X^{q+1}},
\end{align}
where $|\Psi(k)|^2$ is assumed to be even such that $\{b_q(k)\}$ are
alternatively even and odd, leading to $\avg{b_q(k),k^{q+1}} = 0$.
Let the output counts be $n_q^{(\pm)}$.  The mean counts are
\begin{align}
N g_q^{(\pm)} &\approx \frac{N}{2}
\int dX \bk{c_q X^q\pm c_{q+1} X^{q+1}}^2 F(X).
\end{align}
Subtracting one count by the other, the mean is
\begin{align}
N (g_q^{(+)}-g_q^{(-)}) &\approx 2 N c_q c_{q+1} \theta_{2q+1},
\end{align}
so an estimator of the odd moment $\theta_{2q+1}$ can be constructed
as $\check\theta_{2q+1} = (n_q^{(+)}-n_q^{(-)})/(2Nc_qc_{q+1})$.  The
variance of $n_q^{(+)}-n_q^{(-)}$ is
$N (g_q^{(+)}+g_q^{(-)}) \approx
N(c_q^2\theta_{2q}+c_{q+1}^2\theta_{2q+2})$, so the mean-square error
becomes \cite{tsang17,tsang18a}
\begin{align}
  \abbrev{MSE}_{2q+1}^{({\rm SPADE})}  \approx 
\frac{1}{4N}\bk{\frac{\theta_{2q}}{c_{q+1}^2} + \frac{\theta_{2q+2}}{c_{q}^2}}
 = \frac{O(\Delta^{2q})}{N},
  \label{mse_odd}
\end{align}
and the SNR becomes
\begin{align}
\abbrev{SNR}_{2q+1}^{({\rm SPADE})} &\approx
\frac{N(g_q^{(+)}-g_q^{(-)})^2}{g_q^{(+)}+g_q^{(-)}} 
=  N O(\Delta^{2q+2}).
\end{align}
For the first moment ($q = 0$), the error is the same as the well
known $O(1)/N$ error for point-source localization
\cite{farrell66,deschout}.  For the third and higher moments, however,
there is significant enhancement over direct imaging. Note also that
$n_q^{(+)}+n_q^{(-)}$ can give information about the even moments as
well.


\subsection{Fourier object analysis via moments}
The moments can be used in a (generalized) Fourier analysis that may
be more familiar to opticians \cite{villiers}. Suppose that $F(X)$ can
be expanded as
\begin{align}
F(X) &= \sum_{\mu=0}^\infty \tilde F_\mu h_\mu(X)G(X),
\label{fourier}
\end{align}
where $G(X)$ is a nonnegative reference density,
$\{h_\mu(X) = \sum_{\nu=0}^\mu H_{\mu\nu} X^\nu: \mu \in \mathbb
N_0\}$ are orthogonal polynomials that satisfy
\begin{align}
\intall dX G(X) h_\mu(X)h_\nu(X) &= \delta_{\mu\nu},
\end{align}
and $\{\tilde F_\mu\}$ are generalized Fourier coefficients. Each
$h_\mu(X)$ has $\mu$ distinct zeros on the support of $G(X)$
\cite{dunkl}, so each $h_\mu(X)G(X)$ can be regarded as a wavelet that
exhibits localized oscillations.  The Fourier coefficients can be
expressed as
\begin{align}
\tilde F_\mu &= \int dX h_\mu(X)F(X) = \sum_{\nu=0}^\mu H_{\mu\nu} \theta_\nu.
\end{align}
In other words, each Fourier coefficient of order $\mu$ can be
reconstructed from moments up to order $\mu$.  Thus the number of
accurately estimated moments can be regarded as a measure of
resolution, and SPADE can help by bringing in more accurate moments
and increasing the number of obtainable Fourier coefficients for a
subdiffraction object.

With a finite number of moments or Fourier coefficients and no other
prior information, the reconstruction of $F(X)$ is ill-posed and
requires regularization \cite{villiers}. Many linear or nonlinear
algorithms can be used, depending on the application \cite{villiers}.

\subsection{Quantum limits}
Through the Helstrom information, we have learned earlier that SPADE
is optimal for estimating the separation of two point
sources. References~\cite{helstrom70,tsang17} show that direct imaging
is close to optimal for locating a subdiffraction object with a known
shape, while Ref.~\cite{tsang17} also shows that SPADE is close to
optimal for estimating its size. Generalizing such results for
arbitrary moments is much more difficult, as there are now an infinite
number of parameters and an infinite number of spatial modes.  Zhou
and Jiang \cite{zhou19} showed essentially that any measurement should
give a Fisher information that scales with $\Delta$ as
\begin{align}
  \abbrev{FI}_{\mu} &= N O(\Delta^{-\mu_1}),
  &
    \mu_1 &\le \mu,
\label{zhou}
\end{align}
where $\mu_1$ is an integer.  With the Cram\'er-Rao bound
$\abbrev{MSE}_\mu \ge 1/\abbrev{FI}_{\mu}$, the SNR should scale as
\begin{align}
  \abbrev{SNR}_\mu &\le \theta_\mu^2\abbrev{FI}_{\mu} = NO(\Delta^{\mu_2}),
  &
    \mu_2 &\ge \mu,
\end{align}
where $\mu_2$ is another integer. This means that, for a given $\mu$,
the SNR must decrease for smaller $\Delta$, and the decrease is faster
for higher $\mu$. The best scaling with $\Delta$ is achieved at
$\mu_1 = \mu_2 = \mu$, matching the scaling of the SPADE error given
by Eq.~(\ref{mse_even}) for the even moments.  Zhou and Jiang did not
provide a tractable bound on the prefactor of Eq.~(\ref{zhou}),
however, so it remains a question whether SPADE is at all close to the
quantum limit in absolute terms, or there may yet be superior
measurements.

Using more standard quantum estimation theory, Ref.~\cite{tsang19}
proves a quantum limit given by
\begin{align}
\abbrev{FI}_{\mu} \le 
\abbrev{HI}_{\mu} \le \abbrev{HI}'_{\mu} = 
                    NO(\Delta^{-2\lfloor\mu/2\rfloor}),
\label{hi2}
\end{align}
where $\abbrev{HI}'$ is an absolute limit that does not depend on the
measurement and can be approximated analytically or numerically.  The
scaling of $1/\abbrev{HI}'_{\mu}$ with $\Delta$ matches the errors of
SPADE given by Eqs.~(\ref{mse_even}) and (\ref{mse_odd}), suggesting
that SPADE is close to quantum-optimal for both even and odd moments,
but a more quantitative comparison of the quantum limit with the SPADE
performance remains to be done. A limit on the SNR is
\begin{align}
\abbrev{SNR}_\mu \le \theta_\mu^2\abbrev{HI}'_{\mu}
&= NO(\Delta^{2\lceil\mu/2\rceil}).
\end{align}
For a given subdiffraction object, Ref.~\cite{tsang19} also shows that
$\theta_\mu^2\abbrev{HI}'_{\mu}$ must decay quickly with higher $\mu$,
meaning that higher moments are fundamentally more difficult to
estimate.


\section{Other generalizations}
\subsection{Unknown centroid}
A crucial assumption in the preceding discussion is that the object is
highly concentrated near a known coordinate $X = 0$, and the SPADE
device is ideally aligned with $X = 0$. To put it the other way,
$\Delta$ should be regarded as the object width plus any misalignment
of SPADE with the object centroid, and misalignment can reduce the
enhancement by increasing the effective $\Delta$. As direct imaging
can locate the centroid accurately, the misalignment can be minimized
if the object of interest has been imaged before and its centroid is
already known accurately, as is often the case in
astronomy. Otherwise, some overhead photons should be used to locate
the centroid first. Grace and coworkers found that, despite the
overhead, SPADE can still offer significant enhancements of the
two-point resolution over direct imaging with the same total photon
number \cite{grace19}.

In principle, it turns out to be possible to estimate the centroid and
the separation simultaneously at the quantum limit if a multiphoton
measurement is performed, as demonstrated by Parniak and coworkers
\cite{parniak18,chrostowski17}, but the applicability of their
measurement to usual light sources is questionable.

\subsection{Strong thermal light}
While the model of weak thermal light and Poisson statistics works
well for astronomical or fluorescent sources at optical frequencies,
thermal sources at lower frequencies or scattered laser sources can
exhibit super-Poisson statistics \cite{mandel}. Nair computed the
Helstrom information for separation estimation with the exact thermal
state and also proposed variations of SPADE and SLIVER to approach it
\cite{nair_tsang16}. Lupo and Pirandola computed the quantum limit for
the same problem but assumed arbitrary quantum states, including the
thermal state as a special case \cite{lupo}. Yang and coworkers
studied the use of mode homodyne or heterodyne detection for the
two-point problem and found that, although it is not competitive for
weak thermal light, it can offer an enhancement over direct imaging
for strong thermal light \cite{yang17}.

For radio and microwave frequencies, photon shot noise is negligible
at typical temperatures, and heterodyne detection in any spatial-mode
basis is quantum-optimal in the low-frequency limit
\cite[Appendix~A2]{tsang19}. As amplitude measurements via antennas
are already the standard detection method there and they are usually
contaminated with substantial excess noise, the ideas here are not
relevant to those frequencies unfortunately.

\subsection{Two point sources with unequal brightnesses}
\rehacek\ and coworkers studied the quantum limits and the optimal
measurements for two point sources with unequal brightnesses
\cite{rehacek17a,rehacek18}. They found that, while significant
enhancements over direct imaging remain possible, the performance gets
worse for unequal sources.  In hindsight, this is perhaps not
surprising, as moments up to the third are needed to fully parametrize
unequal sources and the SNR for the third moment is fundamentally
poorer. The use of SPLICE for this case was also studied by
Bonsma-Fisher and coworkers \cite{bonsma19}, while the
three-dimensional case was recently studied by Prasad
\cite{prasad19a}.

\subsection{More than two point sources}
Bisketzi and coworkers \cite{bisketzi19} and Lupo, Huang, and Kok
\cite{lupo19} recently proposed methods to compute the quantum limit
to the localization of more than two point sources. Bisketzi and
coworkers found numerically that, regardless of the number of
sources, the Helstrom information matrix retains only two nonzero
eigenvalues as the source separations approach zero. This result is
complementary to---and consistent with---existing results on moment
estimation \cite{tsang19,zhou19}, demonstrating the harsh quantum
limits to imaging beyond centroid and size estimation. As the location
parameters they considered are related nonlinearly to the moment
parameters, the Helstrom information matrix transforms in a nontrivial
way \cite{hayashi}, and a more quantitative comparison of
Ref.~\cite{bisketzi19} with Refs.~\cite{tsang19,zhou19} will require
further effort.

Lupo and coworkers also studied the achievability of the general
quantum limit via interferometers \cite{lupo19}. More work remains to
be done to ascertain whether their proposed interferometer design can
be implemented without knowing the unknown parameters.

\subsection{Excess detector noise}
If the detectors are contaminated with excess noise besides photon
shot noise, the estimation performance necessarily suffers. Len and
coworkers studied the Fisher information of SPADE in the presence of
such noise \cite{len20}, while Lupo studied the quantum limits
\cite{lupo20}.  A fair comparison of these results with noisy direct
imaging remains to be done, however.  Considering that the ideal model
of direct imaging assumes an infinitesimal pixel size, an infinite
number of pixels, no excess noise, and perfect calibration of all
pixels, imperfections in real life may well be even more detrimental
to direct imaging.

\subsection{Partially coherent sources}
Larson and Saleh studied the separation estimation problem for two
partially coherent sources and suggested that Rayleigh's curse would
recur \cite{larson18,larson19}. Their work has been challenged by
Refs.~\cite{tsang_comment19,lee19,wadood19}, however.
Reference~\cite{tsang_comment19} points out a few problems with Larson
and Saleh's analysis, such as the use of a formula for the Helstrom
information that becomes questionable for partially coherent sources.
References~\cite{tsang_comment19,lee19,wadood19} also show that SPADE
can overcome the curse as long as the sources are not highly
correlated, contrary to Larson and Saleh's claim. Another interesting
work on this topic was done by Hradil and coworkers \cite{hradil19},
who also used the questionable formula; see Appendix~\ref{sec_thermal}
for details. In any case, the debate is irrelevant to observational
astronomy and fluorescence microscopy, where there is no sound reason
to doubt the established model of spatially incoherent sources
\cite{goodman_stat,pawley}.

\subsection{Two-dimensional imaging}
Although I have so far focused on imaging in one dimension for
pedagogy, the same principles carry over to two dimensions.  For two
point sources, there are now two parameters for their vectoral
separation.  The quantum limits for the two parameters are the same as
that for the one-dimensional case, and SPADE with respect to the
transverse-electromagnetic (TEM) modes or a pair of SLIVER devices can
still estimate the vectoral separation near the quantum limit
\cite{ant}. For extended sources in two dimensions, a generalization
of the PAD and iPAD modes have been studied in
Refs.~\cite{tsang17,tsang18a,zhou19}, and quantum limits have been
studied in Refs.~\cite{tsang17,zhou19}.

\subsection{\label{sec_3d}Three-dimensional imaging}
Reference~\cite{localization} studies quantum limits to the
three-dimensional localization of one point source as well as two
coherent sources using the full vectoral electromagnetics model (the
discussion of incoherent sources there is flawed and superseded by
Ref.~\cite{tnl}). In the context of the paraxial model on the other
hand, the axial dimension requires special treatment
\cite{goodman}. For the axial localization of one point source,
\rehacek\ and coworkers demonstrated that direct imaging with a
judicious defocus, a common technique in localization microscopy
\cite{diezmann17,zhou19b}, can attain the quantum limit
\cite{rehacek19}. Backlund, Shechtman, and Walsworth computed the
quantum limit to the three-dimensional localization of a point source
using a scalar wave model and proposed special interferometers to
achieve it \cite{backlund18}. Yu and Prasad
\cite{yu18,prasad19,prasad19a} and Napoli and coworkers
\cite{napoli19} studied the same problem but for two incoherent
sources. Zhou and coworkers recently demonstrated a FRFT
interferometer to enhance the estimation of the axial separation
between two sources \cite{zhou19a}.

\subsection{Spectroscopy}
Donohue and coworkers demonstrated mode-selective measurements to
enhance time and frequency estimation for incoherent optical pulses
\cite{donohue18}. On a more mathematical level, the quantum model of a
photon from incoherent sources coincides with that of a quantum probe
subject to random displacements, as pointed out by
Ref.~\cite{tsang19}, so noise spectroscopy with optomechanics or spin
ensembles is another potential application of the theory
\cite{ng16,gefen19}.

\subsection{Biased estimators}
The simplest form of the Cram\'er-Rao bound is applicable to unbiased
estimators only, and it turns out that biased estimators may violate
it significantly \cite{lehmann98}. For example, the Cram\'er-Rao bound
for separation estimation with direct imaging blows up to infinity as
$\theta \to 0$, but the maximum-likelihood estimator, being biased for
this problem, can still achieve a finite error for all $\theta$
\cite{huang11,tham17,tang16}.  For SPADE, the maximum-likelihood
estimator can also violate the Cram\'er-Rao bound and give a vanishing
error as $\theta \to 0$ \cite{tnl}. Given these violations, one may
wonder if the Cram\'er-Rao bound is meaningful outside the theoretical
construct of asymptotic statistics \cite{lehmann98} after all. The
loophole can be fixed by using a Bayesian version of the Cram\'er-Rao
bound \cite{vantrees} that is valid for any biased or unbiased
estimator.
Reference~\cite{tsang18} shows that, from the Bayesian and minimax
perspectives, there remains a significant performance gap between
direct imaging and SPADE for separation estimation, even if biased
estimators are permitted.

\subsection{One-versus-two hypothesis testing}
Another way of defining the two-point resolution is to consider the
error probabilities of deciding whether there is one point source or
two point sources with the same total brightness. As mentioned in
Sec.~\ref{sec_helstrom}, Helstrom performed a pioneering study of this
problem using his quantum detection theory \cite{helstrom73b}, but his
proposed measurement depends on the separation in the two-source
hypothesis, he did not suggest any experimental setup to realize it,
and he did not show how much improvement it could offer.  In the
context of direct imaging, the problem was also studied in
Refs.~\cite{harris64,acuna,shahram04,shahram06}.

Coming in full circle, Lu and coworkers recently showed that the
quantum limit to the hypothesis-testing problem is indeed a
substantial improvement over direct imaging, and both SPADE and SLIVER
can reach the quantum limit in the sub-Rayleigh regime, without
knowing the separation in advance \cite{lu18}.

\section{\label{sec_compare}Comparison 
with other imaging techniques}

In the wider context of imaging research, SPADE is but one of the
countless superresolution proposals in the literature. It nonetheless
possesses many unique advantages and avoids some common pitfalls of
prior ideas, thanks to its firm footing in quantum optics and
statistics.  Its advantages over direct imaging and computational
techniques have already been emphasized in previous sections, and here
I highlight some other important or popular ideas in imaging and how
SPADE compares.

\subsection{Stellar interferometry}
SPADE perhaps bears the most resemblance to stellar interferometry
\cite{goodman_stat,labeyrie,roddier88}, as they are both examples of
applying coherent optical processing to incoherent imaging. In
particular, SLIVER resembles the folding and rotation-shearing
interferometers in optical astronomy, the only difference being that
the former is placed at the image plane and the latter usually at the
pupil plane \cite{roddier88}. Conventional wisdom suggests, however,
that the main advantage of stellar interferometry lies in its
robustness against atmospheric turbulence
\cite{goodman_stat,labeyrie,roddier88}. To quote Goodman
\cite{goodman_stat}: ``The reader may well wonder why the Fizeau
stellar interferometer, which uses only a portion of the telescope
aperture, is in any way preferred to the full telescope aperture in
this task of measuring the angular diameter of a distant object. The
answer lies in the effects of the random spatial and temporal
fluctuations of the earth's atmosphere (`atmospheric seeing')...  It
is easier to detect the vanishing of the contrast of a fringe in the
presence of atmospheric fluctuations than it is to determine the
diameter of an object from its highly blurred image.''  Zmuidzinas
\cite{zmuidzinas03} also suggests that ``the imperfect beam patterns
of sparse-aperture interferometers extract a sensitivity penalty as
compared with filled-aperture telescopes, even after accounting for
the differences in collecting areas.''  No work before ours recognized
that interferometry can outperform direct imaging on statistical terms
for diffraction-limited, filled-aperture telescopes.

Another use of stellar interferometry is to increase the baseline by
coherently combining light from multiple apertures
\cite{labeyrie}. Our theory can also be applied to this multi-aperture
scenario if we take the optical transfer function $\Psi(k)$ defined by
Eq.~(\ref{Psi}) to be the total aperture function for all the
apertures. While conventional interferometer designs call for the
interference of light from pairs of apertures \cite{labeyrie} or the
mimicking of image-formation optics \cite{labeyrie,zmuidzinas03}, our
theory offers the novel insight that demultiplexing the light in terms
of the PAD or iPAD modes associated with $\Psi(k)$ can bring
substantial advantages. This perspective generalizes the recent
studies on the quantum optimality of stellar interferometry
\cite{stellar,pearce17,howard19,lupo19}.

Another idea that sounds similar to SLIVER is nulling interferometry
\cite{labeyrie}, which was proposed for the specific purpose of
exoplanet detection.  The idea there is to remove the light from a
bright star via destructive interference while leaving the light from
a nearby planet intact, but its fundamental statistical performance in
the subdiffraction regime has not been studied to our knowledge.  It
remains open questions whether nulling interferometry or similar ideas
in astronomy turn out to perform similarly to SLIVER or SPADE, and how
the quantum-inspired techniques and the quantum limits may benefit
important astronomical applications in practice, such as exoplanet
detection.

\subsection{Multiphoton coincidence}
While modern stellar interferometers all rely on amplitude
interference \cite{labeyrie}, also called $g^{(1)}$ measurements in
quantum optics, the intensity interferometer by Hanbury Brown and
Twiss---a $g^{(2)}$ measurement---deserves a mention as well,
considering that it inspired the foundation of quantum optics
\cite{mandel} and is still being held in high regard by quantum
opticians. In astronomy, however, the intensity interferometer has in
fact been obsolete for decades because of its poor SNR
\cite{goodman_stat,labeyrie}. It relies on the postselection of
two-photon-coincidence events, which are much rarer than the
one-photon events used in amplitude interferometry and therefore must
give much less information in principle. For example, Davis and Tango
reported an amplitude interferometer that obtained similar results to
those from the intensity interferometer, using only $\sim$2\% of the
observation time \cite{davis86}. For microscopy, the use of
multiphoton coincidence has recently been demonstrated in some heroic
experiments \cite{genovese16,schneider18,tenne19,berchera19}, but
again its statistical performance needs to be studied more
carefully. SPADE, on the other hand, is a $g^{(1)}$ measurement that
relies on the much more abundant one-photon events without the need
for coincidence detection and its statistical performance has been
proved rigorously.

\subsection{Electron microscopy and near-field microscopy}
If the object is on a surface and accessible, then no technique can
compete with electron microscopy, atomic force microscopy, and
scanning-tunneling microscopy in terms of resolution. Those techniques
impose stringent requirements on the sample however, and that is why
optical microscopy remains useful, especially for biological imaging,
as it is able to image biological samples in a more natural
environment and provide protein-specific contrast via fluorophore
tagging.

In terms of optics, near-field techniques have not been successful
because of the short depth of focus and other technical challenges
\cite{betzig15}. In recent years, the use of plasmonics and
metamaterials to enhance the near field \cite{pendry04} has also
attracted immense interest in the academia, but the requirement of
close proximity to the object and the impact of loss remain
showstoppers in practice \cite{khurgin15}.

Being a far-field technique, SPADE is more compatible with biological
imaging, not to mention its unique capability for astronomy and remote
sensing. Unlike metamaterials, SPADE requires only low-loss optical
components and there is no stringent requirement on their feature
size, so fabrication is more straightforward.

Given the theoretical similarity between optical imaging and electron
microscopy \cite{bettens,vanaert}, the application of SPADE to the
latter is possible in principle and indeed tantalizing, but more
research concerning its implementation for electrons needs to be done.

\subsection{Superresolution fluorescence microscopy}

Far-field superresolution techniques such as PALM and STED have been
hugely successful in biological flourescence microscopy
\cite{hell15,moerner15,betzig15}, but many of them rely on
sophisticated control of the source emission, which introduces many
other problems, such as the need for special fluorophores, slow speed
in the case of PALM, and phototoxicity in the case of STED. SPADE, on
the other hand, is a passive far-field measurement that can complement
or supersede the superresolution techniques by extracting more
information from the light or alleviating the need for source
control. The combination of SPADE with microscope configurations, such
as confocal and structured illumination \cite{pawley}, awaits further
research.

\subsection{Nonclassical light}
The application of nonclassical light to sensing and imaging has been
an active research topic in quantum optics for many decades
\cite{kolobov,dowling08,demkowicz15,taylor16,pirandola18,moreau19,fabre19}. It
is now well known, however, that nonclassical light is extremely
fragile against loss and decoherence \cite{demkowicz15,taylor16}, and
any theoretical advantage can be easily lost in practice, not to
mention that the efficient generation and detection of nonclassical
light remain very challenging. More recent proposals, such as quantum
illumination and quantum reading \cite{pirandola18}, apply to
high-noise scenarios, but the achievable improvement turns out to be
quite modest even in theory \cite{tan}.

As SPADE works with classical light, linear optics, and photon
counting, loss and other imperfections are not nearly as detrimental.
If we are to believe that the second quantum revolution is near and
applications using nonclassical resources will soon be widespread
\cite{dowling03}, then SPADE should be an even surer bet.

For astronomy, obviously the light sources cannot be controlled, but
the use of entangled photons and quantum repeaters has been proposed
to teleport photons in stellar interferometry and increase its
baseline \cite{gottesman,khabiboulline19}. Unfortunately, quantum
repeaters are nowhere near practical yet, and conventional linear
optical devices remain the best option in the foreseeable future.

\subsection{Superoscillation, amplification, postselection}

There are so many other superresolution ideas that going through them
all would not be feasible. I list here only a few more:
superoscillation \cite{rogers13}, amplification \cite{kellerer16}, and
postselection \cite{rafsanjani17}. They either require steep
trade-offs with the SNR or have questionable statistics
\cite{prasad94,lantz17}. These examples once again demonstrate the
importance of a rigorous analysis using quantum optics and
statistics. It is important to keep in mind that superresolution is
possible even with direct imaging and data processing, and it is
ultimately limited by the SNR \cite{villiers}. A superresolution
technique is viable only if it can beat direct imaging on statistical
terms.

\section{Conclusion}

Just as the design of engines must go beyond mechanics and consult
thermodynamics, the design of optical sensing and imaging systems must
go beyond electromagnetics and consult statistics. With the
increasingly dominant role of photon shot noise in modern
applications, quantum mechanics is also relevant. Quantum information
theory can tackle all these subjects in one unified formalism, setting
limits to what we can do, and also telling us how much further we can
go.  For incoherent imaging, it gives us the pleasant surprise that
there is still plenty of room for improvement, and we just need to
find a way to achieve it. We found one in the form of SPADE, which
requires only low-loss linear optics and photon counting. While we
started with the simple model of two point sources, we have since
generalized the theory to deal with any subdiffraction object, showing
that substantial improvements remain possible. The theoretical
groundwork has been laid, proof-of-principle experiments have been
done, and applications in astronomy and fluorescence microscopy can
now be envisioned. Special-purpose applications that require only the
low-order moments, such as two-point resolution and object-size
estimation, should be the first to benefit, while more general imaging
protocols will require further research.

Many open problems still remain. On the theoretical side, the exact
quantum limits to general imaging and the optimal measurements to
achieve them remain unclear. The theory for three-dimensional imaging
and spectroscopy remains underdeveloped. On the practical side, an
efficient implementation of SPADE at the right wavelengths is needed
for applications. The performance of SPADE in the presence of
atmospheric turbulence and other technical noises also needs to be
assessed. Fortunately, adaptive optics \cite{esposito11},
photodetectors \cite{michalet13}, and photonics in general have become
so good in recent years that we can be optimistic about reaching the
quantum limits in the near future.

\section*{Acknowledgments}
I am grateful to the seminal contributions of the authors of
Refs.~\cite{tnl,tsang17,tsang18a,dutton19,tsang19,zhou19,tsang19b,sliver,tnl2,nair_tsang16,lupo,tsang18,ant,lu18,rehacek17,yang17,kerviche17,chrostowski17,rehacek17a,rehacek18,backlund18,napoli19,yu18,prasad19,prasad19a,larson18,tsang_comment19,larson19,bonsma19,grace19,bisketzi19,lupo19,lee19,gefen19,hradil19,len20,lupo20,tang16,tham17,paur16,yang16,donohue18,parniak18,paur18,hassett18,zhou19a,paur19,wadood19,rehacek19},
especially the crucial roles of Ranjith Nair, Xiao-Ming Lu, and Shan
Zheng Ang in our early papers. I also acknowledge useful discussions
with Luis S{\'a}nchez-Soto, Jaroslav \rehacek, Zden{\v{e}}k Hradil,
Saikat Guha, and Cosmo Lupo in the course of writing this manuscript.
This work is supported by the Singapore National Research Foundation
under Project No.~QEP-P7.

\appendix
\section{\label{sec_crb}Cram\'er-Rao bound
and Fisher information}
Let
$\{P_Y(y|\theta) > 0: y \in \Omega, \theta \in \Theta \subseteq
\mathbb R\}$ be a family of probability distributions for an observed
random variable $Y$, where $\theta$ is an unknown scalar parameter and
the support $\Omega$ is assumed to be countable and common to all
distributions for simplicity. Let $\check\theta(Y)$ be an estimator of
$\theta$. Define the mean-square error as
\begin{align}
\abbrev{MSE}(\theta) &\equiv \mathbb E\Bk{\check\theta(Y)-\theta}^2
= \sum_{y} P_Y(y|\theta)\Bk{\check\theta(y)-\theta}^2,
\end{align}
where $\mathbb E$ denotes the expectation. The unbiased condition is
\begin{align}
\expect\Bk{\check\theta(Y)} &= \theta.
\end{align}
Under certain regularity conditions on the distributions, the
Cram\'er-Rao bound given by Eq.~(\ref{crb}) holds for any unbiased
estimator, where the Fisher information is \cite{lehmann98}
\begin{align}
\abbrev{FI}(\theta) &\equiv  \sum_{y} 
\frac{1}{P_Y(y|\theta)}\Bk{\parti{P_Y(y|\theta)}{\theta}}^2.
\end{align}
Generalization for probability densities is straightforward
\cite{lehmann98}.

\section{\label{sec_hi}Helstrom information}
Let $\{\rho(\theta): \theta \in \Theta\subseteq \mathbb R\}$ be a family
of density operators for a quantum object. Under a quantum
measurement, the generalized Born's rule is given by
\begin{align}
P_Y(y|\theta) &= \trace E_Y(y)\rho(\theta),
\end{align}
where $\trace$ denotes the operator trace and $E_Y(y)$ is called the
positive operator-valued measure (POVM), which models the measurement
statistics \cite{hayashi}.  Define the Helstrom information as
\cite{helstrom}
\begin{align}
\abbrev{HI} &= \trace \rho L^2 = \trace \parti{\rho}{\theta} L,
\label{hidef}
\end{align}
where $L$ is a solution to
\begin{align}
\parti{\rho}{\theta} &= \frac{1}{2} \bk{\rho L+L\rho}.
\label{sld}
\end{align}
For any POVM, Helstrom proved $\abbrev{MSE} \ge \abbrev{HI}^{-1}$
\cite{helstrom}, while Nagaoka \cite{nagaoka89} and Braunstein and
Caves \cite{braunstein} proved
\begin{align}
\abbrev{FI}(\theta) &\le \abbrev{HI}(\theta).
\label{hibound}
\end{align}
Although they also proved that
$\max_{E_Y}\abbrev{FI}(\theta) = \abbrev{HI}(\theta)$ and a projection in
the eigenstates of $L$ gives an optimal POVM, it is important to keep
in mind that $L$ is a function of $\theta$, and the optimal POVM
derived from it at one value of $\theta$ may be suboptimal at other
values.  In practice, obviously $\theta$ is unknown, and there is no
guarantee that one can find a POVM that is optimal across a range of
$\theta$.  A solution, proposed by Nagaoka and refined by Hayashi and
Matsumoto \cite{hayashi05} and Fujiwara \cite{fujiwara2006}, is to
consider repeated adaptive measurements, and they showed that the
total Fisher information of such measurements can approach the
Helstrom information in the limit of infinitely many measurements
under certain technical conditions.

\section{\label{sec_thermal}Thermal state 
in the ultraviolet limit}
Consider thermal light in one temporal mode and multiple spatial
modes, and let $\{a_0,a_1,\dots\}$ be the annihilation operators for
the spatial modes.  As first proposed by Glauber \cite{glauber06}, the
thermal state is \cite{helstrom}
\begin{align}
\sigma &= \expect\bk{\ket{\alpha}\bra{\alpha}}
= \int d^2\alpha \Phi(\alpha)\ket{\alpha}\bra{\alpha},
\label{glauber}
\\
\Phi(\alpha) &= 
\frac{1}{\det(\pi\Gamma)}\exp\bk{-\alpha^\dagger\Gamma^{-1}\alpha},
\label{thermal}
\end{align}
where $\alpha = (\alpha_0,\alpha_1,\dots)^\top$ is a column vector of
zero-mean complex Gaussian random variables with probability density
$\Phi$, $\top$ denotes the transpose, $\dagger$ denotes the conjugate
transpose, $\ket{\alpha}$ is a multimode coherent state that obeys
$a_q\ket{\alpha} = \alpha_q\ket{\alpha}$, and $\Gamma$ is the mutual
coherence matrix \cite{mandel}. In particular, the first moments of
$\alpha$ are given by
\begin{align}
\mathbb E(\alpha) &= 0,
&
\mathbb E(\alpha\alpha^\top) &= 0,
&
\mathbb E(\alpha\alpha^\dagger) &= \Gamma.
\end{align}
The photon-counting distribution is
\begin{align}
P(n) &= \bra{n}\sigma\ket{n} = \mathbb E\abs{\braket{n|\alpha}}^2,
\label{mandel}
\\
\ket{n} &= \prod_q \frac{(a_q^{\dagger})^{n_q}}{\sqrt{n_q!}}
\ket{{\rm vac}},
\\
\abs{\braket{n|\alpha}}^2&= 
\exp(-\alpha^\dagger\alpha)\prod_q \frac{|\alpha_q|^{2n_q}}{n_q!},
\end{align}
where $\ket{n}$ is a Fock state and $\ket{{\rm vac}}$ is the vacuum
state.  Equation~(\ref{mandel}) agrees with the semiclassical theory
by Mandel \cite{mandel}. With $M$ temporal modes, the density operator
can be modeled as $M$ copies of $\sigma$, or
\begin{align}
\rho &= \sigma^{\otimes M}.
\label{tensor}
\end{align}

To simplify the thermal state for optical frequencies, let
\begin{align}
  \epsilon &\equiv \trace\Gamma
\end{align}
be the average photon number per temporal mode and
\begin{align}
    g &\equiv \frac{\Gamma}{\trace\Gamma}
\end{align}
be the normalized mutual coherence matrix. Define the ultraviolet
limit as $\epsilon \to 0$ while holding $N = M\epsilon$ constant.  The
zero-photon probability per temporal mode is
\begin{align}
P(0,\dots,0) &= \mathbb E\Bk{\exp(-\alpha^\dagger\alpha)} 
= 1-\epsilon + O(\epsilon^2),
\end{align}
the one-photon probability is
\begin{align}
P(0,\dots,n_q=1,0,\dots)
&= \mathbb E\Bk{\exp(-\alpha^\dagger\alpha)|\alpha_q|^2}
\nonumber\\
&= \epsilon g_{q} + O(\epsilon^2),
\end{align}
where the diagonal entries of a matrix are abbreviated as
$g_{qq} = g_q$, and the probability of two or more photons is
$O(\epsilon^2)$.  The photon counts summed over $M$ temporal modes
hence become Poisson in the ultraviolet limit \cite{goodman_stat}.  A
simplified quantum model in this limit is \cite{stellar,tnl}
\begin{align}
\sigma &= (1-\epsilon)\ket{{\rm vac}}\bra{{\rm vac}}
+ \epsilon \rho_1 + O(\epsilon^2),
\label{weak_sigma}
\end{align}
where the one-photon density operator is 
\begin{align}
\rho_1 &= \sum_{q,p} g_{qp}\ket{\phi_q}\bra{\phi_p},
&
\ket{\phi_q} &= a_q^\dagger\ket{{\rm vac}}.
\label{rho1}
\end{align}
For paraxial incoherent imaging in particular \cite{tsang17},
\begin{align}
  \rho_1
  &= \intall dX F(X)e^{-i\hat kX}\ket{\psi}\bra{\psi}e^{i\hat kX},
\label{rho1imaging}
\end{align}
where $\hat k$ is the spatial-frequency or momentum operator,
$\ket{\psi}$ is the one-photon state with spatial wavefunction
$\braket{x|\psi} = \psi(x)$, and $\ket{x}$ is the one-photon position
eigenket that obeys $\braket{x|x'} = \delta(x-x')$.
$f(x) = \bra{x}\rho_1\ket{x}$ gives Eq.~(\ref{direct}), while
$g_q = \bra{\phi_q}\rho_1\ket{\phi_q}$ gives Eq.~(\ref{Pq}).  If $f$
and $g$ depend on $\theta$ (but $\epsilon$ does not), the Fisher
information for the Poisson processes is given by
Eqs.~(\ref{fi_direct2}) and (\ref{fi}).

The ultraviolet limit and the negligence of $O(\epsilon^2)$ terms mean
that multiphoton coincidence events and bunching effects are ignored
\cite{goodman_stat}.  Besides thermal sources, the model here also
applies to any incoherent sources, such as fluorescent sources
\cite{pawley} or even electrons \cite{bettens,vanaert}, as long as
they obey an incoherent-imaging model with Poisson counting
statistics.

For the thermal state given by Eqs.~(\ref{glauber}) and
(\ref{thermal}), Helstrom showed that \cite{helstrom}
\begin{align}
\abbrev{HI} &= \trace\parti{\Gamma}{\theta}\Upsilon,
\label{hithermal}
\end{align}
where $\Upsilon$ is a solution to
\begin{align}
\parti{\Gamma}{\theta} &= 
\frac{1}{2}
\Bk{\Gamma\Upsilon(I+\Gamma)+(I+\Gamma)\Upsilon\Gamma},
\label{hld}
\end{align}
and $I$ is the identity matrix.  Reference~\cite[Appendix~A]{tsang19}
shows that the information given by Eqs.~(\ref{hithermal}) and
(\ref{hld}) on a per-photon basis is upper-bounded by its ultraviolet
limit, which coincides with the information computed for the
one-photon density operator $\rho_1$ given by Eq.~(\ref{rho1}) if
$\epsilon$ does not depend on $\theta$, viz.,
\begin{align}
\frac{\abbrev{HI}^{(\sigma)}}{\epsilon} &\le 
\lim_{\epsilon\to 0}\frac{\abbrev{HI}^{(\sigma)}}{\epsilon} =
\abbrev{HI}^{(\rho_1)}.
\label{hiweak}
\end{align}
With $M$ temporal modes, the Helstrom bound is multiplied by $M$
\cite{hayashi}, so $\abbrev{HI}^{(\rho)} = M\abbrev{HI}^{(\sigma)}$,
and the total information in the ultraviolet limit becomes
\begin{align}
\abbrev{HI}^{(\rho)} &\le 
\lim_{\epsilon\to 0}\abbrev{HI}^{(\rho)} = N \abbrev{HI}^{(\rho_1)},
\end{align}
which means that $\abbrev{HI}^{(\rho_1)}$ also serves as a limit for
thermal states with arbitrary $\epsilon$ if $\epsilon$ does not depend
on $\theta$.

If $\epsilon$ depends on $\theta$, which may happen with partially
coherent sources \cite{tsang_comment19}, one must be more careful and
go back to Eqs.~(\ref{hithermal}) and (\ref{hld}). For
$\epsilon \ll 1$, $I + \Gamma \approx I$, and Eq.~(\ref{hld}) can be
approximated as
\begin{align}
\parti{\Gamma}{\theta} &\approx
\frac{1}{2}
\bk{\Gamma\Upsilon+\Upsilon\Gamma}.
\label{hld2}
\end{align}
Equations~(\ref{hithermal}) and (\ref{hld2}), in terms of the mutual
coherence matrix $\Gamma$, resemble Eqs.~(\ref{hidef}) and (\ref{sld})
in terms of the density operator $\rho$. Notice, however, that
Eqs.~(\ref{hithermal}) and (\ref{hld2}) are in terms of the
\emph{unnormalized} $\Gamma$. References~\cite{larson18,hradil19}, on
the other hand, use the normalized version $g = \Gamma/\trace\Gamma$
in the formulas and may have produced unphysical results for partially
coherent sources.

\section{\label{sec_gs}Gram-Schmidt process}
Consider an inner-product space equipped with an inner product
$\avg{u,v}$ between two elements $u$ and $v$ and a norm
$\norm{u} = \sqrt{\avg{u,u}}$.  An illustrative example is the space
of Euclidean vectors in $\mathbb R^d$, with the dot product as the
inner product and the vector length as the norm. Given a set of
linearly independent elements $S = \{u_0,u_1,\dots\}$, the
Gram-Schmidt process produces an orthonormal basis $\{b_0,b_1,\dots\}$
for the space spanned by $S$ \cite{debnath05}.  The process starts
with
\begin{align}
v_0 &= u_0,
&
b_0 &= \frac{v_0}{\norm{v_0}}.
\end{align}
Then, for each $q = 1,2,\dots$,
\begin{align}
v_q &= u_q - \sum_{p=0}^{q-1} \Avg{u_q,b_p}b_p,
&
b_q &= \frac{v_q}{\norm{v_q}}.
\end{align}
$\norm{b_q} = \sqrt{\avg{b_q,b_q}} = 1$ by design. One can check that
$v_q$ and $b_q$ are orthogonal to $\{b_0,\dots,b_{q-1}\}$. It follows
that $\{b_0,\dots,b_q\}$ is an orthonormal basis with
\begin{align}
\Avg{b_q,b_p} &= \delta_{qp}.
\end{align}
Since the space spanned by $\{b_0,\dots,b_{q-1}\}$ is the same as the
space spanned by $\{u_0,\dots,u_{q-1}\}$, each $b_q$ is also
orthogonal to $\{u_0,\dots,u_{q-1}\}$.

\section{\label{sec_multi}Multiparameter estimation}
Now suppose that $\theta \in \Theta \subseteq\mathbb R^K$ is a column
vector of parameters, and the estimator is also a vector. Define the
mean-square error covariance matrix as
\begin{align}
\abbrev{MSE}_{\mu\nu}(\theta) &\equiv 
\mathbb E\Bk{\check\theta_\mu(Y)-\theta_\mu}
\Bk{\check\theta_\nu(Y)-\theta_\nu}.
\end{align}
Diagonal entries of a matrix are again abbreviated as
$\abbrev{MSE}_{\mu\mu} = \abbrev{MSE}_\mu$.  The multiparameter
Cram\'er-Rao bound \cite{lehmann98} can be expressed as the matrix
inequality 
\begin{align}
\abbrev{MSE} &\ge \abbrev{CRB}
\equiv \abbrev{FI}^{-1},
\\
\abbrev{FI}_{\mu\nu}(\theta) &\equiv
\sum_{y} \frac{1}{P_Y(y|\theta)}
\parti{P_Y(y|\theta)}{\theta_\mu}\parti{P_Y(y|\theta)}{\theta_\nu}.
\end{align}
The matrix inequality means that $\abbrev{MSE}-\abbrev{CRB}$ is
positive-semidefinite \cite{horn}, or equivalently
$u^\top (\abbrev{MSE}-\abbrev{CRB}) u \ge 0$ for any real column
vector $u$. For example, the multiparameter Cram\'er-Rao bounds for
two point sources and more general objects measured with direct
imaging and SPADE have been derived in
Refs.~\cite{tnl,ant,tsang17,tsang18a,tsang19b}.

The Helstrom information matrix is defined as
\begin{align}
\abbrev{HI}_{\mu\nu} &\equiv 
\real \trace \rho L_\mu L_\nu 
=\trace \parti{\rho}{\theta_\mu}L_\nu ,
\\
\parti{\rho}{\theta_\mu} &= 
\frac{1}{2}\bk{\rho L_\mu+L_\mu\rho}.
\end{align}
The matrices can be shown to inherit all the properties of their
scalar version by substituting the directional derivative
$\partial/\partial\theta = \sum_\mu u_\mu\partial/\partial\theta_\mu$
and $L = \sum_\mu u_\mu L_\mu$ for an arbitrary real vector $u$.  For
example, upon the substitutions, the scalar Fisher information becomes
$u^\top\abbrev{FI}u$ and the scalar Helstrom information becomes
\begin{align}
\trace \rho L^2 = u^\top\mathcal H u &= u^\top\abbrev{HI}u,
&
\mathcal H_{\mu\nu} &\equiv \trace \rho L_\mu L_\nu ,
\end{align}
where I have used the fact that, since $u^\top\mathcal H u$ and $u$
are real,
$u^\top\mathcal H u = \real(\sum_\mu u_\mu \mathcal H_{\mu\nu} u_\nu)
= \sum_\mu u_\mu \real(\mathcal H_{\mu\nu}) u_\nu = \sum_\mu u_\mu
\abbrev{HI}_{\mu\nu} u_\nu$. The Nagaoka bound given by
Eq.~(\ref{hibound}) becomes
$u^\top \abbrev{FI} u \le u^\top\abbrev{HI}u$, meaning that
Eq.~(\ref{hibound}) still holds as a matrix inequality. A consequence
of the matrix inequality is that the inverses obey the reverse
relation \cite{horn}, so the Nagaoka bound leads to
\begin{align}
\abbrev{MSE} &\ge \abbrev{FI}^{-1} \ge \abbrev{HI}^{-1}.
\end{align}

\bibliography{contemporary_physics2}

\end{document}